\documentclass[apj, letterpaper]{emulateapj}  % this is for apjlookalike-Showman
\usepackage{apjfonts}  % I used this for apj emulate -- Showman

\usepackage{amsmath}
\newcommand{\m}{\rm \,m}
\newcommand{\K}{\rm \,K}
\newcommand{\J}{\rm \,J}
\newcommand{\kg}{\rm \,kg}
\newcommand{\Pa}{\rm \, Pa}
\newcommand{\W}{\rm \, W}

\newcommand{\cm}{\rm \, cm}
\newcommand{\Hz}{\rm \, Hz}
\newcommand{\erg}{\rm \, erg}

\newcommand{\km}{\rm \, km}

\begin{document}

\slugcomment{\bf}
\slugcomment{Submitted to the Astrophysical Journal}

\title{Atmospheric circulation of hot Jupiters: Three-dimensional
circulation models of HD209458b and HD189733b with simplified forcing} 

\shorttitle{Atmospheric Circulation of Hot Jupiters}
\shortauthors{Showman et al.}

\author{Adam P.\ Showman\altaffilmark{1}, Curtis S.\ Cooper\altaffilmark{1,2},
Jonathan J.\ Fortney\altaffilmark{3,4,5}, and Mark S. Marley\altaffilmark{5}}

\altaffiltext{1}{Department of Planetary Sciences and Lunar and Planetary
Laboratory, The University of Arizona, 1629 University Blvd., Tucson, AZ 85721 USA; showman@lpl.arizona.edu}
\altaffiltext{2}{NASA Astrobiology Institute, University of Arizona, Tucson, AZ 85721}
\altaffiltext{3}{UCO/Lick Observatory, Department of Astronomy \& Astrophysics, University of California, Santa Cruz, CA 95064}
\altaffiltext{4}{Spitzer Fellow}
\altaffiltext{5}{NASA Ames Research Center 245-3, Moffett Field, CA}

% Abstract 
\begin{abstract}
\label{Abstract}

We present global, three-dimensional numerical simulations of
the atmospheric circulation on HD209458b and HD189733b and
calculate the infrared spectra and light curves predicted
by these simulations, which we compare with available observations.
Radiative heating/cooling is parameterized with
a simplified Newtonian relaxation scheme.  Our simulations develop 
day-night temperature contrasts that vary strongly with pressure.
At low pressure ($<10\,$mbar), 
air flows from the substellar point toward the antistellar point,
both along the equator and over the poles.  At deeper levels,
the flow develops an eastward equatorial jet with speeds
of 3--$4\km\sec^{-1}$, with weaker westward flows at high latitudes.
This basic flow pattern is robust to variations in model resolution,
gravity, radiative time constant, and initial temperature structure. 
Nightside spectra show deep
absorption bands of H$_2$O, CO, and/or CH$_4$, whereas on
the dayside these absorption bands flatten out or even flip
into emission.  This results from the strong effect of dynamics
on the vertical temperature-pressure structure; the temperature 
decreases strongly with altitude on the nightside but becomes 
almost isothermal on the dayside.  In Spitzer
bandpasses, our predicted planet-to-star flux ratios vary
by a factor of $\sim2$--10 with orbital phase, depending
on the wavelength and chemistry.  
For HD189733b, where a detailed 8-$\mu$m light curve has 
been obtained, we correctly produce the observed phase 
offset of the flux maximum, but we do not explain the 
flux minimum and we overpredict the total flux variation.
This discrepancy likely results from the simplifications
inherent in the Newtonian relaxation scheme and provides motivation
for incorporating realistic radiative transfer in future
studies.

\end{abstract}


%%%%%%%%%%%%%%%%%%%%%%%
% Begin document body %
%%%%%%%%%%%%%%%%%%%%%%%

\section{Introduction}
\label{Introduction}

Of the $\sim255$ extrasolar giant planets (EGPs) that have been 
discovered to date (see www.exoplanet.eu), the hot Jupiters
have proved the easiest to characterize because of their high 
temperatures, short orbital periods, and likelihood of transiting 
their stars \citep{charbonneau-etal-2007}.  
The planetary mass and radius can be calculated from radial-velocity and 
transit data, allowing the surface gravity to be inferred; furthermore,
the rapid spindown times for such close-in planets \citep{guillot-etal-1996} 
suggest that these planets are tidally locked, providing an estimate
of their rotation rates.  Such tidal locking implies that, in the
absence of atmospheric winds, the dayside would be extremely hot
and the nightside would be extremely cold.

Impressively, infrared light curves have been obtained with the 
Spitzer Space Telescope for at least six hot Jupiters, helping
to constrain the three-dimensional (3D) temperature pattern and
atmospheric circulation of these objects.  Continuous 8-$\mu$m
observations of HD189733b over half an orbit demonstrate that the planet's
photospheric temperatures are relatively homogenized: the nightside 
is perhaps only $\sim200$--$300\K$ colder than the dayside 
\citep{knutson-etal-2007b}.  The exquisite temporal resolution
allows detailed structure to be inferred.  The light curve shows
that the flux peaked $16\pm6^{\circ}$ of orbital phase
before secondary eclipse; a flux map of the planet
constructed from the light curve suggests that the hottest region
lies $\sim30^{\circ}$ longitude east of the substellar point and 
the coldest region lies a comparable amount west of the antistellar
point, providing clear evidence for advection by winds 
\citep{knutson-etal-2007b}.  8-$\mu$m light curves for 
HD209458b and 51 Peg b have also been obtained \citep{cowan-etal-2007}
and similarly imply significant thermal homogenization by winds,
although these light curves contain insufficient temporal sampling
to determine displacements (if any) of the hottest and coldest regions 
from the substellar and antistellar points. 
In contrast, Ups And b and HD 179949 
exhibit large-amplitude phase variations, with no observationally
significant phase shift, which suggests that these 
planets have large day-night temperature differences (perhaps $>500\K$)
that track the stellar insolation patterns \citep{harrington-etal-2006,
cowan-etal-2007}.  Together, these observations suggest that the 
degree to which winds homogenize the temperatures varies from planet 
to planet.

There exist many additional observational constraints on the circulation, 
including dayside spectra, upper limits on albedo, and constraints 
on composition.  For HD209458b, Spitzer secondary-eclipse photometry 
at wavelengths of 3.6--$24\,\mu$m  suggests the existence of a hot
stratosphere on the dayside 
\citep{knutson-etal-2007d, burrows-etal-2007b}, while 
HD189733b seems to lack such a temperature inversion.  Spitzer 
IRS spectra of these planets' daysides from 7--$14\,\mu$m lack obvious 
molecular absorption bands \citep{grillmair-etal-2007, 
richardson-etal-2003, richardson-etal-2007, swain-etal-2007},
although debate exists about the interpretation 
\citep{fortney-marley-2007}.   Transit spectroscopy observations 
have yielded detections of sodium \citep{charbonneau-etal-2002, 
redfield-etal-2008} and water \citep{barman-2007, tinetti-etal-2007b}
on both planets, and upper limits on
several other compounds.

Several groups have investigated the atmospheric circulation on
hot Jupiters \citep[for a review of the approaches 
see][]{showman-etal-2007}.  
\citet{cho-etal-2003, cho-etal-2008} and 
\citet{langton-laughlin-2007, langton-laughlin-2008} 
solved variants of the two-dimensional 
(2D) or quasi-2D equations on a rotating sphere.  The
advantage of this approach is that, by reducing the vertical 
resolution to one layer, 
simulations with high horizontal resolution can be 
performed. The idealizations implicit in such simplified 
models also allow the model dynamics to be relatively easily understood;
a comparison of such results with 3D models and
observations can lend important insights into which dynamical
processes cause which outcomes \citep{vasavada-showman-2005,
showman-etal-2007}. This has proved important, for example, in studying
horizontal vortex/jet interactions on Solar-System giant planets 
and in Earth's stratosphere \citep{cho-polvani-1996a, showman-2007, 
polvani-etal-1995}.  

On the other hand, one-layer models 
do not represent the vertical structure of the flow, and they 
exclude inherently 3D processes such as baroclinic 
instabilities\footnote{Baroclinic instabilities 
are a form of sloping convection that can occur in the 
presence of horizontal temperature contrasts in rotating,
statically stable atmospheres.  Lateral and vertical motion 
forces the cold air downward and the hot air upward,
which releases potential energy that helps to drive the
circulation.  Most large-scale winter weather in the United States,
Europe, and Asia is caused by baroclinic instabilities.},
vortex tilting, and vertical wave propagation.
In some contexts, such as the tropospheres of the Earth, Mars,
Jupiter, and Saturn, 3D processes 
(e.g., baroclinic instabilities and convection) are crucial
for determining the mean state because they 
convert potential energy to kinetic energy, hence causing
the accelerations that pump the east-west jet streams. 
To some degree, such 3D processes can be parameterized in 2D
models, but this introduces uncertainties regarding
(for example) the rate at which the parameterized processes
inject energy into the flow.
In the hot-Jupiter context, a 3D approach is further
motivated by the expectation that the radiative 
time constant varies by orders of magnitude in the vertical 
\citep{iro-etal-2005}, which in the presence of the day-night
heating gradient would cause patterns of temperature and wind that
vary both vertically and horizontally in an inherently 3D manner. 
Because infrared spectra of planetary atmospheres depend sensitively
on the vertical temperature profile, a full 3D representation
of the winds and temperatures is also necessary for robustly 
predicting infrared spectra and wavelength-dependent light curves
in Spitzer bandpasses. 

In light of the above, we have adopted a 3D approach to
investigate the atmospheric circulation of hot Jupiters
\citep{cooper-showman-2005, cooper-showman-2006, 
showman-guillot-2002}.
\citet{cooper-showman-2005, cooper-showman-2006} 
and \citet{showman-guillot-2002} focused on HD209458b and
approximated the dayside heating and nightside cooling
using a Newtonian cooling/heating
scheme, which parameterizes the radiative heating rate (in $\K\sec^{-1}$)
as $(T_{\rm eq} - T)/\tau_{\rm rad}$, where $T_{\rm eq}$ is the
specified radiative-equilibrium temperature profile (hot on the
dayside, cold on the nightside), $T$ is the actual temperature,
and $\tau_{\rm rad}$ is the radiative-equilibrium timescale,
which was taken to be
a function of pressure.  In \citet{cooper-showman-2005},
the vertical structure of $T_{\rm eq}$
and $\tau_{\rm rad}$ were taken from \citet{iro-etal-2005}. 
The simulations showed development of several broad jets --- 
including a superrotating\footnote{Superrotation refers to eastward
zonal winds, that is, winds moving faster than the planetary rotation.}
equatorial jet with speeds up to several $\km\sec^{-1}$ ---
and exhibited horizontal (day-night) temperature differences
that varied strongly in height. \citet{cooper-showman-2006}
extended these simulations to include carbon chemistry,
which showed that interconversion between CO and CH$_4$ should
become chemically quenched at low pressure, leading
to nearly constant abundances of these species everywhere
above the photosphere at abundances that depend
on the temperature in the deep ($\sim10\,$bar) atmosphere.
Using a multi-stream plane-parallel radiative-transfer code, 
\citet{fortney-etal-2006b} calculated theoretical infrared light curves 
and spectra for HD209458b from the 3D temperature patterns of 
\citet{cooper-showman-2005, cooper-showman-2006}.
These calculations suggested that the day-night flux differences
could reach factors of $\sim2$--10 depending on wavelength
and that the peak IR emission should lead the
secondary eclipse by $\sim2$--3 hours in most Spitzer IRAC bands.

Here, we present new global, 3D numerical simulations, spectra,
and light curves that continue the research program begun by 
\citet{cooper-showman-2005, cooper-showman-2006} and
\citet{fortney-etal-2006b}.  Major
improvements are incorporated in several areas.  First,
\citet{cooper-showman-2005, cooper-showman-2006} treated
the day-night difference in $T_{\rm eq}$ as a free parameter
(with values ranging from 100--$1000\K$) rather than the
result of a laterally varying radiative-transfer calculation.  
This uncertainty in $\Delta T_{\rm eq}$ directly translates
into uncertainty in the light curves and spectra.  Within the
context of the Newtonian-cooling framework, the best approach ---
which we take here --- is to explicitly calculate the 
longitude, latitude, and height-dependent radiative-equilibrium 
temperatures for use in the Newtonian heating/cooling scheme.
$\Delta T_{\rm eq}$ is thus no longer a free parameter but
represents the true difference in radiative-equilibrium temperature
from the dayside to the nightside.  Likewise, we here
calculate $\tau_{\rm rad}$ over the full 3D grid and express
this as a function of {\it both} temperature
and pressure rather than treating it as a function of $p$ alone 
(based on the 1D global average temperature profile) as done by 
\citet{cooper-showman-2005, cooper-showman-2006}.

Second, we consider not only HD209458b (the canonical planet emphasized
by most previous circulation studies) but HD189733b, and perform
parameter variations in planetary rotation rate, gravity, and
radiative time constant to determine how these parameters affect
the behavior.  Finally, we calculate infrared spectra and light curves
from the 3D temperature patterns for comparison with available
observations, including the recent 8-$\mu$m light curve for
HD189733b \citep{knutson-etal-2007b}, secondary-eclipse 
photometry for HD189733b \citep{deming-etal-2006}, and
secondary-eclipse photometry for HD209458b, which now spans 
all available Spitzer channels \citep{knutson-etal-2007d,
deming-etal-2005a}.

A long-term goal is to couple the dynamics to a realistic
representation of radiative transfer.  However, this is a difficult
task, and experience with planetary general circulation models 
(GCMs) shows that the radiative-transfer calculation can
become a computational bottleneck, making such a model computationally
expensive. The computational and conceptual simplicity of Newtonian 
cooling provide compelling arguments for documenting the extent
to which a model driven by such simplified forcing can explain
available observations.  The results described herein thus
provide a benchmark documenting the best that can be done 
with this simplified approach.

In \S 2 we describe the dynamical model and present
the calculations of laterally varying radiative-equilibrium temperature
and radiative time constant.  \S 3 presents the basic dynamical
results and parameter variations.  \S 4 describes the spectra
and infrared light curves in Spitzer bandpasses as predicted by
the simulations. \S 5 compares our results with pertinent aspects
of dynamical calculations by other authors. \S 6 shows that
the global-scale flow of hot Jupiters should be close to
local hydrostatic balance, and \S 7 concludes.

\section{Model}

\subsection{Dynamics}

We solve the global, three-dimensional primitive equations in spherical
geometry using the ARIES/GEOS dynamical core \citep{suarez-takacs-1995}.
The primitive equations are the standard equations for large-scale
flow in stably stratified atmospheres whose horizontal dimensions
greatly exceed the vertical dimensions.  This is expected to be
true on hot Jupiters, which have horizontal scales of $10^7$--$10^8\m$
but atmospheric scale heights of only 200--$500\km$, leading
to a horizontal:vertical aspect ratio of $\sim20$--500.  This large
aspect ratio allows the vertical momentum equation to be replaced
with local hydrostatic balance, meaning that the local vertical
pressure gradient $\partial p/\partial z$ is balanced by the
local fluid weight $\rho g$ 
\citep[for a derivation see, e.g.,][]{holton-2004}. The primitive equations 
admit the full range of balanced motions, buoyancy (gravity) waves, 
rotationally modified (e.g., Kelvin and Rossby) waves, and horizontally 
propagating sound waves, but they filter vertically propagating sound 
waves. The horizontal momentum, vertical momentum, mass continuity, 
and thermodynamic energy equations are as follows \citep[e.g.,][p.~60--67]
{kalnay-2003}:

\begin{equation}
{d{\bf v}\over dt}= -\nabla \Phi - f {\bf k}\times {\bf v}
\label{momentum}
\end{equation}
\begin{equation}
{\partial \Phi\over \partial p}=-{1\over\rho}
\label{hydrostatic}
\end{equation}
\begin{equation}
\nabla \cdot {\bf v} + {\partial \omega\over\partial p}=0
\label{continuity}
\end{equation}
\begin{equation}
{d T\over dt} = {q\over c_p} + {\omega \over \rho c_p}
\label{energy}
\end{equation}
where ${\bf v}$ is the horizontal velocity on constant-pressure surfaces, 
$\omega\equiv dp/dt$ is the vertical velocity in pressure coordinates, 
$\Phi$ is the gravitational potential on constant-pressure surfaces, 
$f\equiv 2\Omega\sin\phi$ is the
Coriolis parameter, $\Omega$ is the planetary rotation rate
($2\pi$ over the rotation period), ${\bf k}$ is the
local vertical unit vector, $q$ is the thermodynamic heating rate
($\W\kg^{-1}$), and $T$, $\rho$, and $c_p$ are the
temperature, density, and specific heat at constant pressure.
$\nabla$ is the horizontal gradient
evaluated on constant-pressure surfaces, and $d/dt=\partial/\partial t
+ {\bf v}\cdot \nabla + \omega \partial/\partial p$ is the material
derivative.  Curvature terms are included in ${\bf v}\cdot\nabla {\bf v}$.
The dependent variables ${\bf v}$, $\omega$, $\Phi$,
$\rho$, and $T$ are functions of longitude $\lambda$, latitude $\phi$,
pressure $p$, and time $t$.  

Note that the vertical velocity is nonzero --- it enters via the mass-continuity 
equation and the thermodynamic energy equation. For a known atmospheric state 
at a given timestep, Eqs.~\ref{momentum} and \ref{energy} are integrated
forward, leading to expressions for ${\bf v}$ over the 3D grid at the subsequent 
timestep.  The horizontal divergence of these velocities is generally nonzero, and the
vertical velocity $\omega$ is then evaluated for that timestep via Eq.~\ref{continuity}
with use of the boundary conditions.  For conditions relevant to hot
Jupiters, the characteristic vertical velocity is typically 
$\sim10\m\sec^{-1}$ near the photosphere 
\citep[][and see \S 6]{showman-guillot-2002, cooper-showman-2006}.

The ARIES/GEOS dynamical core discretizes the equations 
in longitude and latitude using an Arakawa C grid \citep{arakawa-lamb-1977}
and adopts a pressure coordinate in the vertical.  To maintain numerical
stability, we follow standard practice and apply 
Shapiro and polar filtering to the time tendencies
\citep{shapiro-1970, suarez-takacs-1995}.
The top boundary condition is constant pressure and the bottom
boundary condition is an impermeable surface, which we place
far below the region of interest.  These boundaries are free-slip
in horizontal velocity.
We solve the equations using horizontal resolutions in longitude
and latitude of $72\times45$ or $144\times 90$ with 30--40 layers
spaced evenly in $\log p$.  The ARIES/GEOS model has been widely
used in Earth and Mars studies and has been successfully benchmarked 
against standard test cases \citep{held-suarez-1994}.

For HD209458b, we adopt gravity, planetary radius, and
rotation rate $\Omega$ of $9.81\m\sec^{-2}$, $9.44\times10^7\m$,
and $2.06\times10^{-5}\sec^{-1}$ (implying a rotation period of 3.5 days), 
respectively.
The corresponding values for HD189733b are $22.62\m\sec^{-2}$,
 $8.22\times10^7\m$, and $3.29\times10^{-5}\sec^{-1}$ (implying 
a rotation period of 2.2 days), respectively.
For both planets, $c_p = 1.23\times10^4\J\kg^{-1}\K^{-1}$ and the
molar mass is $2.36\times10^{-3}\kg\,{\rm mol}^{-1}$, implying
a specific gas constant of $3523\J\kg^{-1}\K^{-1}$.   We adopt the ideal
gas equation of state.  

The top layer is placed at $0.7\,$mbar. Radiative calculations 
that match the planetary radius suggest that the temperature
profiles converge to the interior adiabat at $\sim50$
bars on HD209458b and $\sim500\,$bars on HD189733b (Fig.~\ref{t-eq}).  
Accordingly, we place the bottom pressure at $100\,$bars for 
HD209458b and $900\,$bars for HD189733b.  The intial conditions
contain no winds and adopt a temperature profile resulting
from a 1D global-average radiative-transfer calculation of
the planet.  Our basic results are insensitive to the initial
condition, as is discussed more fully in \S 3.

The heating rate is
\begin{equation}
{q\over c_p}={T_{\rm eq}(\lambda, \phi, p)-T(\lambda,\phi,p,t)
\over \tau_{\rm rad}(p, T)}
\label{heating}
\end{equation}
We assume that the obliquity and orbital eccentricity are zero,
and that the planet rotates synchronously, which implies that
the radiative-equilibrium temperatures $T_{\rm eq}(\lambda,\phi, p)$ 
and radiative time constants $\tau_{\rm rad}(p, T)$ are time-independent; 
the following subsection describes how we calculate them.  The
substellar latitude and longitude remain fixed at $0^{\circ}$, 
$0^{\circ}$ throughout the simulation.

\subsection{Radiative-equilibrium temperature structures}
 
We specify $T_{\rm eq}$ in Eq.~\ref{heating} by calculating the
radiative-equilibrium temperatures as a function of longitude,
latitude, and pressure for conditions relevant to HD209458b and
HD189733b.  To do so, we calculate atmospheric
pressure-temperature ($p$-$T$) profiles using the plane-parallel
multi-stream radiative transfer code first used for Titan by
\citet{mckay-etal-1989} and later extended to giant planets
and brown dwarfs by Marley, Fortney, and collaborators
\citep{marley-etal-1996, burrows-etal-1997, marley-mckay-1999, 
marley-etal-1999, fortney-etal-2005, fortney-etal-2006a, fortney-etal-2006b}.
Although the metallicities of HD209458b and HD189733b are unknown,
the host stars have near-solar metallicity.  Here we adopt
solar metallicity for the planets and assume local chemical 
equilibrium \citep{lodders-fegley-2002,lodders-fegley-2006}.  
The calculations use a large and continually updated 
opacity database described in \citet{freedman-etal-2007}.
For these exploratory calculations we neglect cloud opacity, 
although the ``rainout'' of elements that condense into clouds 
is always properly accounted for.

For the 
present calculations, the opacities of TiO and VO are excluded, 
as has been standard for models of multi-Gyr-old hot Jupiters at the 
incident flux levels expected for gas giants 
beyond $\sim0.03\,$AU (for a Sun-like primary).  One-dimensional 
atmosphere models suggest that gasous TiO and VO are cold trapped at 
pressures of $\sim10-100\,$bars where solid clouds incorporating Ti 
and V form, which suggests that TiO and VO should be absent in the 
observable atmosphere \citep[e.g.,][]{hubeny-etal-2003, fortney-etal-2006a}.
However, the recent detection of a hot dayside stratosphere on 
HD209458b \citep{knutson-etal-2007d} suggests that future 
studies of this planet should include 
TiO and VO \citep{burrows-etal-2007b, fortney-etal-2007b}.  
The limited data for HD 189733b are consistent with no TiO/VO opacity 
\citep{fortney-marley-2007}.

The code treats thermal radiation from the planet from
0.26--$325\,\mu$m and incident radiation from the parent star from
0.26--$6.0\,\mu$m.  The calculation is performed at 60 model layers 
from 0.3 mbar to 1000 bars, although the opacities at $p>100\,$bars
are quite uncertain owing to imperfect knowledge of the line
broadening under these conditions.  The calculation includes 
incident flux from the parent star, thermal flux 
from the planet's atmosphere, and thermal flux from the planet's interior, 
parameterized by $T_{\rm int}$, the intrinsic effective temperature.

To calculate radiative-equilibrium $p$-$T$ profiles, we take
an approach similar to that of \citet{barman-etal-2005}.  At a given 
location on the planet's day side, the
incident stellar flux arrives with an angle $\theta$ from the local 
vertical.  The cosine of this angle, $\mu$, varies from 1 at the 
substellar point to 0 at the terminator.  Large
$\mu$ allows deeper penetration of stellar flux, greater absorption, 
and a warmer atmosphere.  Smaller
$\mu$ gives a longer path length and airmass to a given pressure, 
implying shallower penetration of flux, less absorption (greater 
scattering), and a cooler atmosphere.  We calculate
$p$-$T$ profiles for concentric rings of atmosphere that are symmetric 
around the normal at the subsolar point.  The profiles are calculated 
at $\mu=1.0,$ 0.9, 0.8, 0.7, 0.6, 0.5, 0.4, 0.3,
0.2, 0.1, 0.05, 0.02, 0.01, and 0.0.  The latter case, which
is the radiative-equilibrium profile for an isolated object 
with the planet's known radius, gives the
radiative-equilibrium profile across the entire nightside.  
All cases use a constant specific entropy at the bottom boundary, 
corresponding to the value needed to reproduce the planetary radius 
with an assumed heavy element abundance of $\sim$30-40 Earth masses 
in the planetary interior, similar to that of Jupiter 
\citep{saumon-guillot-2004}.

Figs.~\ref{t-eq}a and b depicts the resulting 
profiles for HD209458b and HD189733b, respectively.  These 
atmospheric $p$-$T$ profiles show the atmosphere structure that each planet 
would attain under radiative equilibrium.  The large temperature 
gradients implied on the planetary day side shows that vigorous 
dynamics is expected to occur.  In chemical equilibrium,
CO would dominate over CH$_4$ across much of the dayside,
but CH$_4$ would dominate on the nightside.
In the absence of any energy
redistribution, the night sides of these planets would 
be quite cold, with 100-mbar temperatures of 500 K and 200 K for 
HD209458b and HD189733b, respectively.  The night sides abundances would be 
dominated by H$_2$O, CH$_4$, and NH$_3$, 
and would be devoid of atomic alkalis, which would condense into clouds.  
Perhaps H$_2$O cloud condensation would occur as well.  Dynamical 
redistribution of energy will of course alter the atmospheric 
temperature structure and chemistry.

\subsection{Radiative timescales}

\begin{figure}
\includegraphics[scale=0.55, angle=0]{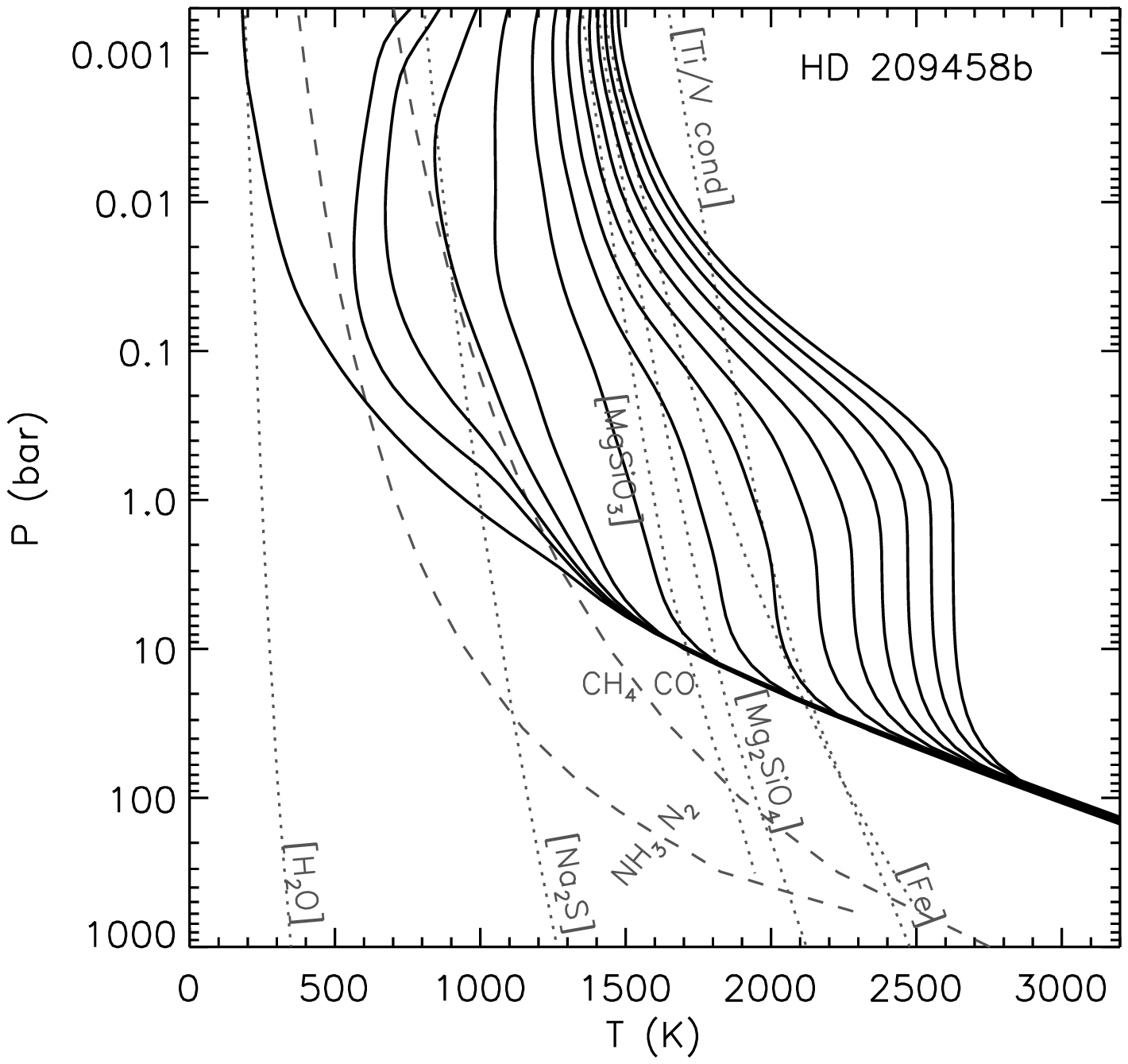}
\includegraphics[scale=0.55, angle=0]{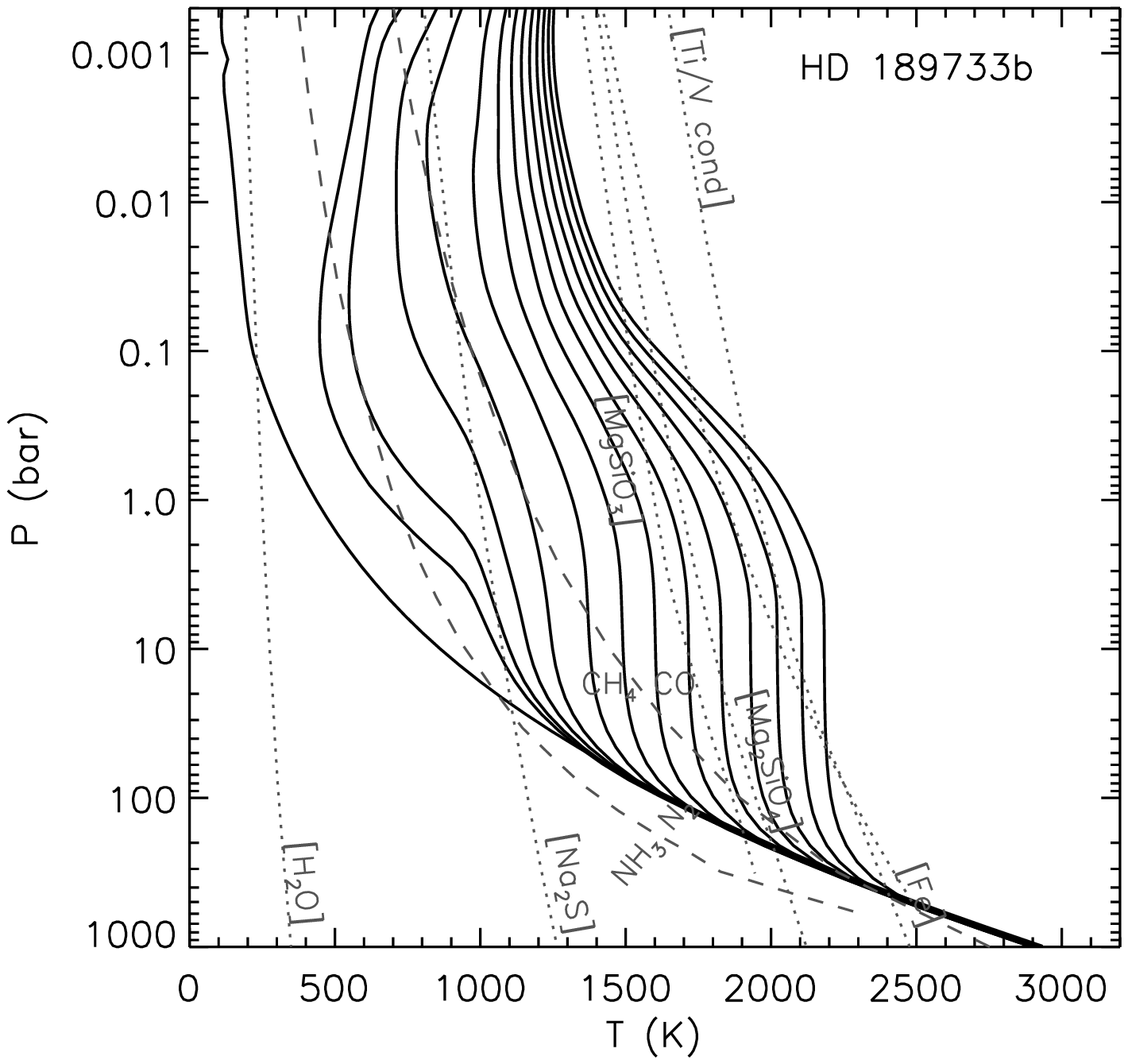}
\caption{
Pressure-temperature (\emph{p-T}) profiles for planet HD 209458b (top) and HD189733b 
(bottom) as a function of $\mu$.  The hottest profile is $\mu$=1.0, 
and the profiles decrease in steps of 0.1 to $\mu$=0.1.  Profiles for 
$\mu$=0.05, 0.02, 0.01 are also shown, as well as the non-irradiated night side 
profile ($\mu=0$).  Condensation curves are shown as dotted lines and are labeled.  
Equal abundance curves for CO/CH$_4$ and N$_2$/NH$_3$ are dashed.
}
\label{t-eq}
\end{figure}

We developed a simple method for calculating the temperature and 
pressure dependent radiative timescales across the large $p$-$T$ 
space accessed by these atmospheres.  We first calculate a suite 
of converged radiative-convective model atmospheres similar those 
shown in Fig.~\ref{t-eq}.
Since we are interested in radiative time 
constants as a function $p$ and $T$ in these atmsopheres, which may in principal be found anywhere in the atmosphere, 
at any value of $\mu$, we calculated $\sim$15 profiles across the 
$p$-$T$ space shown in these figures, but for each profiles assumed 
day-side average conditions, meaning $\mu$=0.5.  We increased and 
decreased the incident stellar flux to reach temperatures from 
$\sim$100-2500 K for both planets.

For each one of these profiles, we add a thermal perturbation 
$\delta T$=10 K to a given model layer with pressure $p_O$.  
Because this perturbed profile deviates from radiative equilibrium, 
the perturbed layer emits a different flux than it receives, implying a 
cooling/heating that should cause the thermal perturbation to decay.  
Given the perturbed profile, we perform a radiative-transfer 
calculation to calculate the net flux versus
height.  For any function $f(t)$, a characteristic timescale
for variation of $f$ is $f (df/dt)^{-1}$ (for exponential decay
this would simply give the e-folding timescale).  Thus, the 
characteristic timescale for decay of the thermal perturbation is $\delta T$ divided by the heating rate expressed
in $\K\sec^{-1}$, namely

\begin{equation}
\tau_{\rm rad}= \delta T {\rho c_p\over dF/dz}
\end{equation}
where $F$ is the net vertical radiative flux, $\rho$ is density,
$z$ is height, and
$dF/dz$ is the volumetric heating rate ($\W\m^{-3}$).  
By varying $p_0$, we can calculate the $\tau_{\rm rad}$ as a function 
of pressure for every $p-T$ profile in our suite of models.  Tests with 
different values of $\delta T$ and those utilizing perturbations that span
multiple model layers yield similar results.

\begin{figure}
\includegraphics[scale=0.55, angle=0]{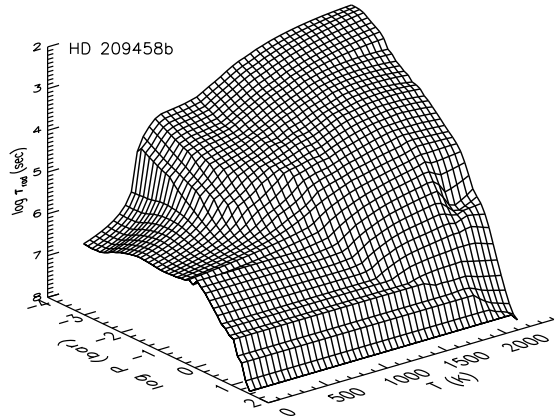}
\includegraphics[scale=0.55, angle=0]{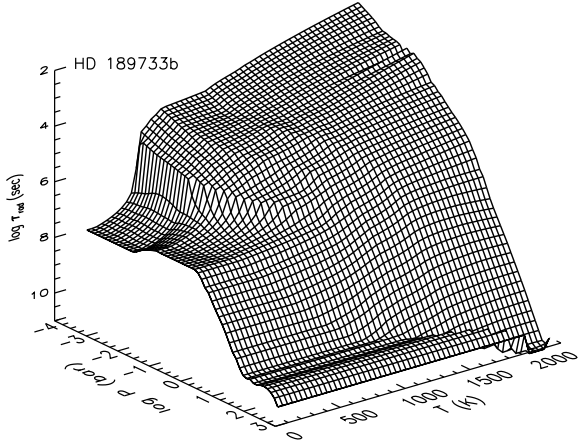}
\caption{Radiative time constant ($\tau_{\rm rad}$) as function of pressure 
and temperature in the atmosphere of HD 209458b (top) and HD189733b (bottom).  
Time constants were derived from planet-wide average \emph{p-T} profiles 
computed with the incident stellar flux increased and decreased to map out 
high and low temperatures.  Small-scale bumpy structure in the 
surfaces is not physically significant. (See text.)}
\label{tau-rad}
\end{figure}

Fig.~\ref{tau-rad} shows the resulting $\tau_{\rm rad}$ for HD209458b 
and HD189733b.  These are surfaces on a regular $p$-$T$ grid, 
interpolated from the original calculations.  Some modest extrapolation 
was done on the high $T$/low $p$ and low $T$/high $p$ 
corners.  Perhaps the most striking aspect of the calculations is the 
strong pressure dependence, with $\tau_{\rm rad}$ values changing by 
8 orders of magnitude at a give temperature, from millibars to hundreds 
of bars in pressure.  
Temperature effects are also substantial.  At millibar pressures we find a 
six order of magnitude change in $\tau_{\rm rad}$ across the temperatures that 
we explore, with the most significant temperature dependence coming below 
$\sim$500 K, where $\tau_{\rm rad}$ becomes quite long.  The calculations 
for HD 189733b extend to higher pressures than HD 209458b because of the 
expected deeper radiative zone in the atmosphere of HD 189733b.  On
average, the values for HD 209458b slightly exceed those for HD 189733b
(by a factor of 1.5--3).  This results from the lower gravity for
HD 209458b, which leads to a ``puffed up'' atmosphere with greater
mass per area across a given pressure interval.

\begin{figure}
\includegraphics[scale=0.5, angle=0]{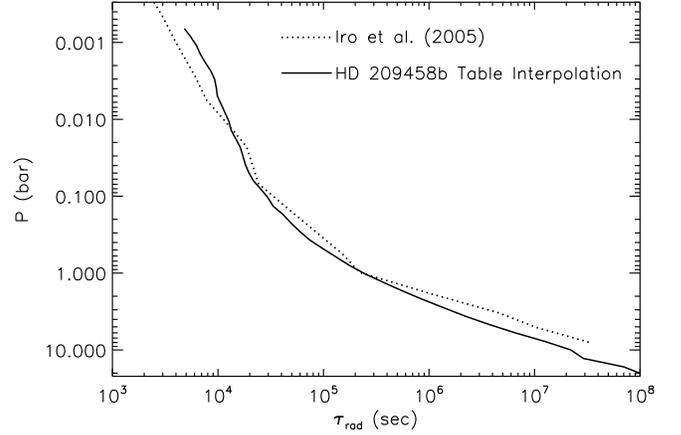}
\caption{
A comparison of the radiative time constant ($\tau_{\rm rad}$) as function 
of pressure for the atmospheric $p$-$T$ profile of HD 209458b from 
\citet{iro-etal-2005}.  The dotted curve is the calculation from \citet{iro-etal-2005}.  
The solid curve is an interpolation (using this same profile) in the 
\emph{p-T}-$\tau_{\rm rad}$ surface for HD 209458b shown in 
Fig.~\ref{tau-rad}.
}
\label{tau-compare}
\end{figure}

To our knowledge the only previous calculation of radiative time constants 
over a wide range in atmospheric pressure for hot Jupiters 
was by \citet{iro-etal-2005} for HD 209458b.  These calculations were limited 
to $\tau_{\rm rad}$ as a function of $p$ only along the 1D atmospheric 
$p$-$T$ profile that these authors calculated for the planet.  
\citet{iro-etal-2005} determined $\tau_{\rm rad}$ by adding Gaussian 
temperature perturbations to their previously converged radiative 
equilibrium profile.  They then performed a time-dependent
radiative-transfer calculation to explicitly simulate the 
characteristic time needed for their model atmosphere to relax 
back to the radiative equilibrium structure.  Our 
\emph{p-T}-$\tau_{\rm rad}$ surfaces shown in Fig.~\ref{tau-rad}
cover a much larger phase space than was explored by these authors.  Although 
our chosen methods are quite different, Fig.~\ref{tau-compare} shows that our 
derived $\tau_{\rm rad}$ values for the 1D profile from \citet{iro-etal-2005} 
are actually very similar.  Tabulated values of $\tau_{\rm rad}$ as a function 
of $p$ and $T$ for planets HD 209458b and HD 189733b are found in 
Tables 1 and 2.  Nevertheless, the precise $\tau_{\rm rad}$ values should
be viewed as uncertain because deviations from radiative equilibrium in
the real atmosphere will not follow the idealized shape adopted 
here.

\medskip
%%%%%%%%%%%%%%%%%%
% FIGURE 1
%%%%%%%%%%%%%%%%%
% note: for emulateapj version I used scale=0.55, but this
% made panels side by side in submission version.  So submitted 
% version used scale=0.7.  These files were originally 
% called pt_cs07_HD209458.eps and pt_cs07_HD189733.eps.

%%%%%%%%%%%%%%
% FIGURE 2
%%%%%%%%%%%%%%
% These were originally called grid209.eps and grid189.eps.
% note: used scale=0.55 for emulateapj version, 0.65 for submitted
% version.

%%%%%%%%%%%%%%%
% FIGURE 3
%%%%%%%%%%%%%%%
% This was originally called Iro_compare.eps.

\section{Results: Dynamical simulations}

\subsection{Flow regime and dependence on parameters}

\begin{figure}
\includegraphics
% Note: to make this, I used ggv to view it, expanded it to 185% of
% its normal size, and then did screen capture using xv, and then
% resaved it as ps using 8-bit and xv's ``compress'' option.
[scale=0.58, angle=180]
{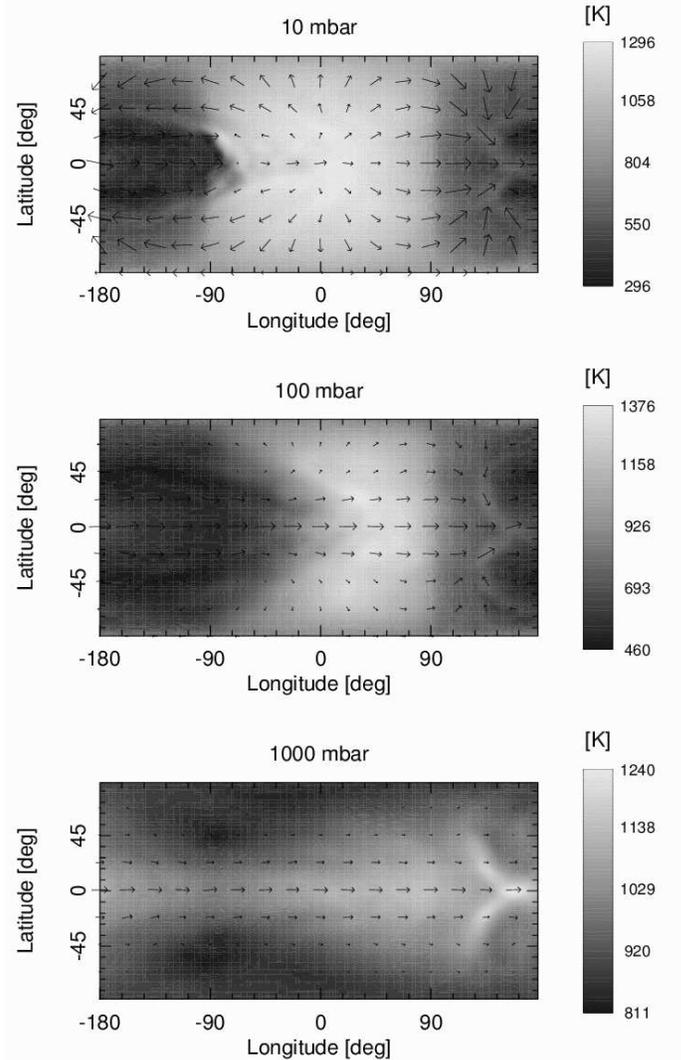}
\caption{Temperature (greyscale) and winds (arrows) for nominal 
HD189733b simulation at 900 days.  Resolution is $144\times90$.  
For this and all simulations in this paper, the substellar point 
is at longitude, latitude ($0^{\circ}$, $0^{\circ}$); the terminators 
are at longitudes $\pm90^{\circ}$.}
\label{189733b-t-winds}
\end{figure}

\begin{figure}
\includegraphics[scale=0.58, angle=180]
{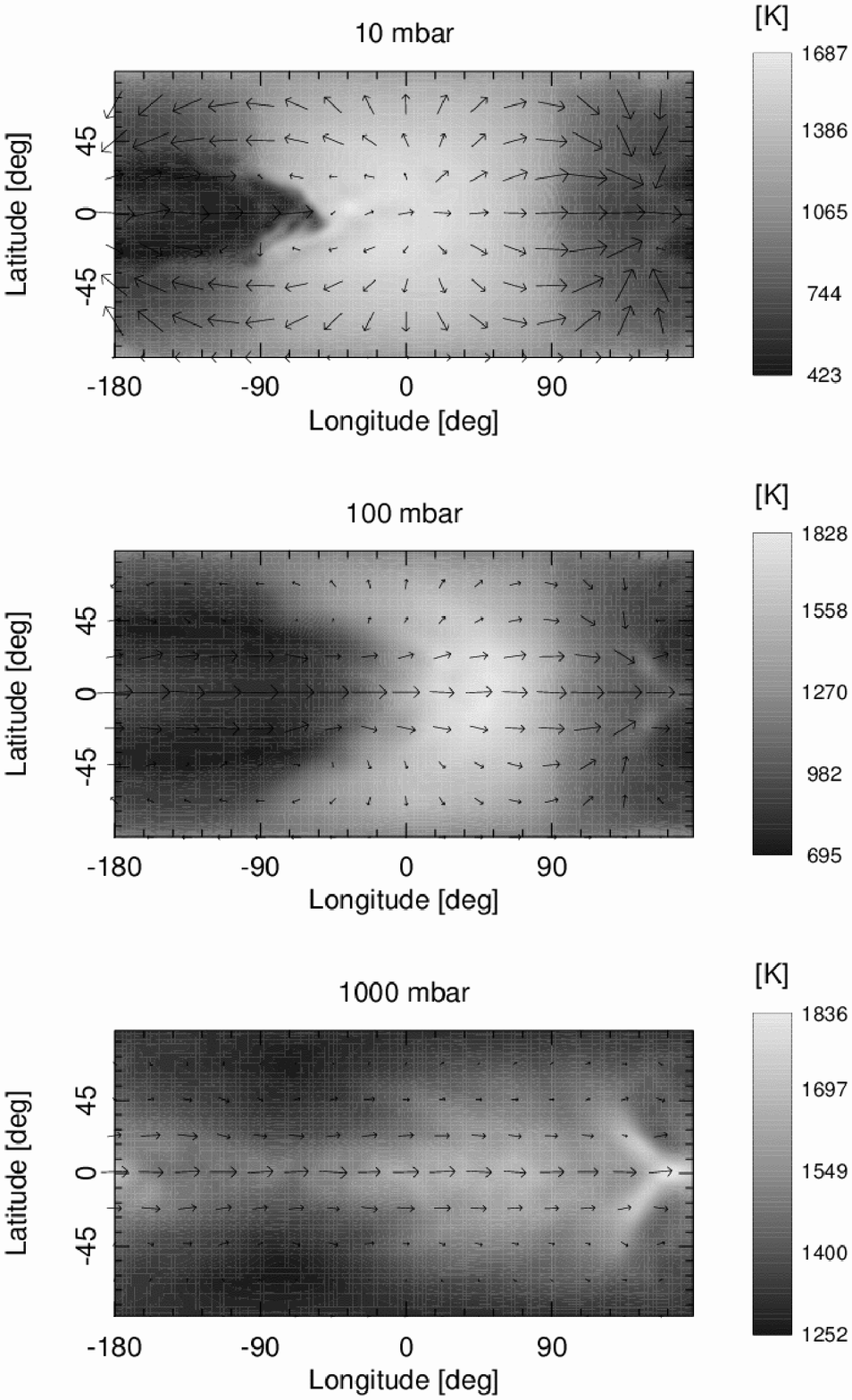}
\caption{Temperature (greyscale) and winds (arrows) for HD209458b 
simulation at 800 days.  Resolution is $144\times90$.}
\label{209458b-t-winds}
\end{figure}

Consistent with the results of \citet{cooper-showman-2005, cooper-showman-2006},
the imposed day-night heating contrasts lead to rapid development of large 
day-night temperature differences and winds reaching several $\km\sec^{-1}$. 
Figures~\ref{189733b-t-winds}--\ref{209458b-t-winds} show the 
temperature (greyscale) and horizontal winds (arrows) for our 
nominal simulations of HD189733b and HD209458b at times
of 900 and 800 days, respectively, by which time the temperature
and wind patterns have reached a quasi-steady state at $p<1\,$bar.
Panels are shown at pressures of 
10 mbar, 100 mbar, and 1 bar, which bracket the range of pressures 
expected to be important for observable spectra and light curves.  
Both simulations use a resolution of $144\times90$ with 30 layers.  

As expected, the flow develops patterns of temperature and wind that 
vary horizontally and vertically in an inherently 3D manner. 
At low pressure ($p\le10\,$mbar), the radiative time constants are 
shorter than the typical
time for wind to advect across a hemisphere, and so the temperature 
patterns track the stellar
heating --- hot on the dayside and cold on the nightside, with 
approximate boundaries
at the terminators (longitudes $\pm90^{\circ}$).  
The predominant flow pattern moves air away from the substellar point 
toward the antistellar point, both along the equator and over the poles.  
At deeper levels,
the flow forms a banded structure dominated by an eastward equatorial 
jet extending from latitude 
$\sim30^{\circ}$N to $30^{\circ}$S 
(Figs.~\ref{189733b-t-winds}--\ref{209458b-t-winds}b and c).   
In the range $\sim50$--$300\,$mbar, the advection times are similar 
to the radiative times, leading to 
large longitudinal temperature differences whose patterns are distorted from 
the day-night heating patterns; 
at $p\ge 1\,$bar, however, the radiative timescale is longer than the 
longitudinal advection time and 
temperatures become homogenized in longitude.  The advection times 
in latitude exceed those in longitude, 
so nonzero latitudinal temperature differences persist even at 
relatively deep levels 
(Fig.~\ref{189733b-t-winds}c and \ref{209458b-t-winds}c).  This banded 
structure results directly
from the effects of planetary rotation, as has been shown by previous 
authors \citep{showman-etal-2007, 
showman-guillot-2002, cho-etal-2003, cho-etal-2008, menou-etal-2003, 
cooper-showman-2005, dobbs-dixon-lin-2008}.

%%%%%%%%%%%%%%%
% FIGURE 4
%%%%%%%%%%%%%%%

Consistent with the results of previous studies \citep{showman-guillot-2002,
menou-etal-2003, cho-etal-2003, cho-etal-2008, langton-laughlin-2007,
dobbs-dixon-lin-2008}, the horizontal lengthscales of the simulated circulation
patterns are comparable to a planetary radius.  This results from the 
fact that the Rhines length and the Rossby deformation radius, 
which tend to control the horizontal dimensions of the dominant flow 
structures, are comparable to a planetary radius for the modest rotation 
rates, high temperatures, and large static stabilities
relevant here \citep{showman-guillot-2002, menou-etal-2003, showman-etal-2007}.

%%%%%%%%%%%%%%%%%%
% FIGURE 5
%%%%%%%%%%%%%%%%%

The {\it qualitative} patterns of wind and temperature are 
similar in our HD189733b and HD209458b simulations, 
suggesting that these planets may have comparable circulation 
regimes despite modest (factor of 1.5--2) differences in gravity, 
rotation rate, and stellar flux.  However, examination of
Figs.~\ref{189733b-t-winds} and \ref{209458b-t-winds} shows that the 
{\it absolute} temperatures differ substantially between these cases.
Our HD209458b simulation produce minimum and maximum temperatures of
$\sim400\K$ and $\sim1700$--$1800\K$, respectively, in the 
observationally relevant layer from 10--$1000\,$mbar.  In contrast, 
our HD189733b simulation is cooler, with temperatures ranging from 
$300\K$ to only $\sim1400\K$ in this same layer.
Although HD189733b lies closer to its parent star (0.0313 AU as compared
to 0.046 AU for HD209458b), the stellar luminosity is lower,
explaining the cooler planetary temperatures.

%%%%%%%%%%%%%%%%%%%%%
% FIGURE 6
%%%%%%%%%%%%%%%%%%%%%
% This was originally called hd209458b-high-res-900d-tav.eps.
\begin{figure}
\includegraphics[scale=0.55, angle=0]
{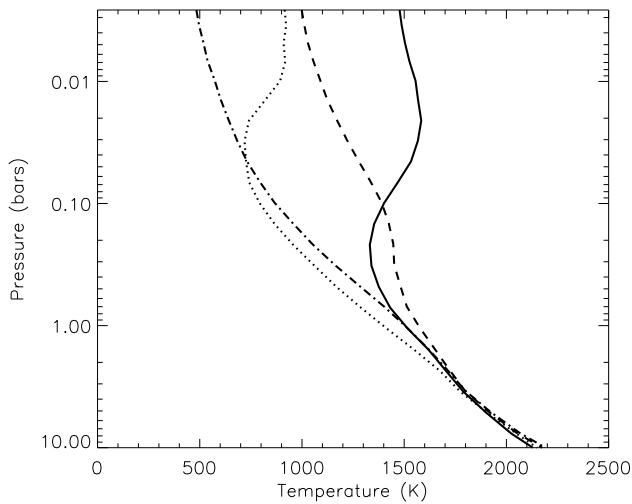}
\caption{Temperature profiles for nominal HD209458b case
averaged across a disk $45^{\circ}$ in radius
centered at the substellar point (solid), $90^{\circ}$ longitude (dashed),
antistellar point (dashed-dotted), and $-90^{\circ}$ longitude (dotted).}
\label{t-profiles}
\end{figure}

A careful comparison of Fig.~\ref{t-eq} with
Figs.~\ref{189733b-t-winds}--\ref{209458b-t-winds} shows how the actual 
temperatures compare to the radiative
equilibrium temperatures.  For HD189733b, at low pressure ($\sim10\,$mbar), 
the local radiative-equilibrium temperature $T_{\rm eq}$ 
ranges from 155--$1340\K$ while the actual temperature ranges
from $\sim300$--$1300\K$ (fluctuating slightly in time).  For HD209458b,
at this same pressure, $T_{\rm eq}$ ranges from 290--$1670\K$ while the
actual temperature ranges from 423--$1687\K$.  Both these comparisons
show that $T$ lies very close to $T_{\rm eq}$ on the dayside, but that
$T$ is substantially warmer than $T_{\rm eq}$ on the nightside.  This
effect results directly from the temperature dependence of $\tau_{\rm rad}$.
At deeper levels, where $\tau_{\rm rad}$ is greater, 
$T$ is substantially warmer than $T_{\rm eq}$
on the nightside and cooler than $T_{\rm eq}$ on the dayside.

%%%%%%%%%%%%%%%%%%%
% FIGURE 7
%%%%%%%%%%%%%%%%%%
\begin{figure}
\includegraphics[scale=0.64, angle=180]
{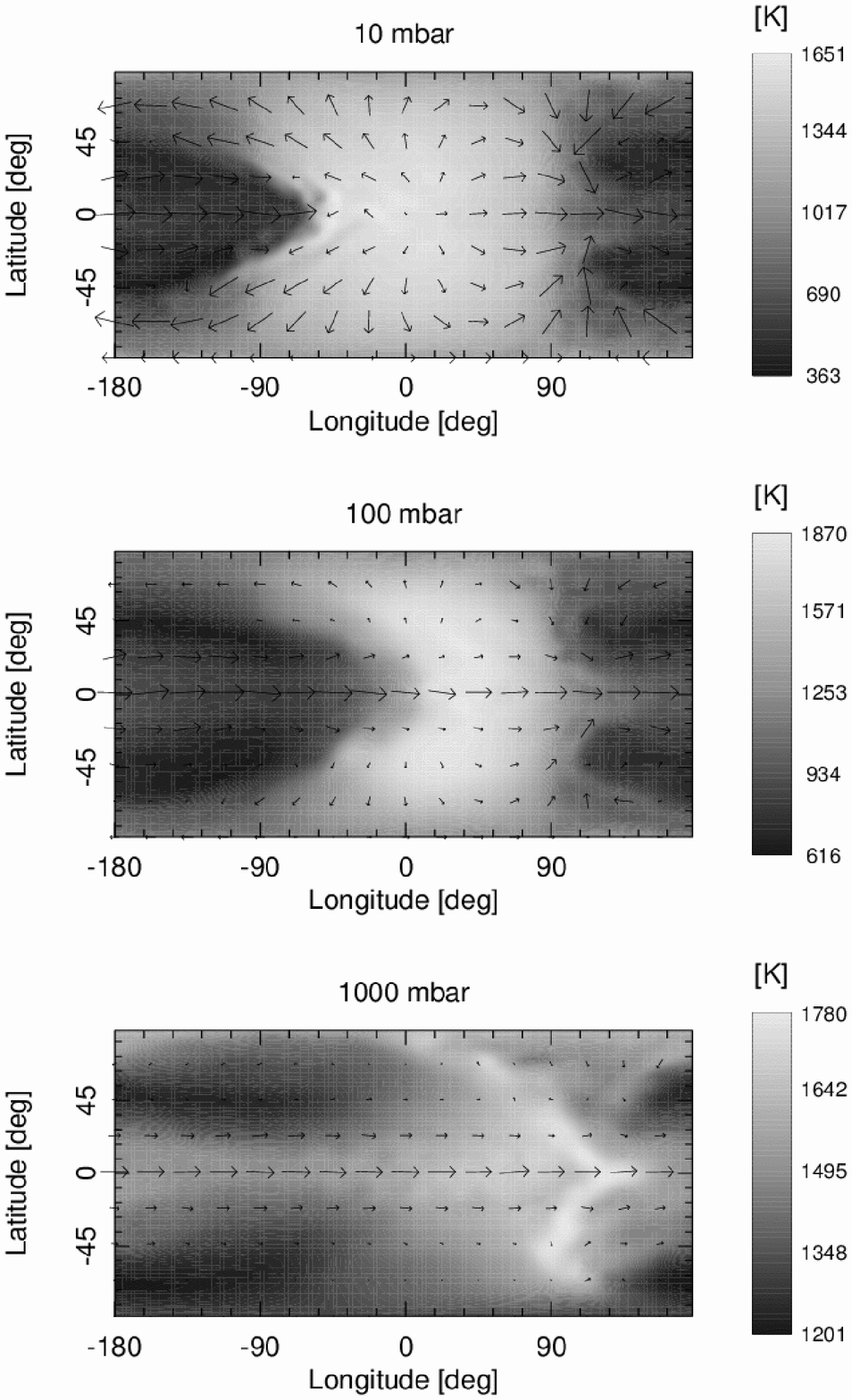}
\caption{Temperature (greyscale) and winds (arrows) for HD209458b 
simulation with twice the nominal $\Omega$ at 800 days. Resolution 
is $144\times90$.}
\label{209458b-double-omega}
\end{figure}

Figure~\ref{t-profiles} shows temperature profiles from our
nominal HD209458b case. To better represent the broad-scale conditions
relevant for spectra and lightcurves (to be discussed in \S 4),
the profiles have been averaged across a circular region $45^{\circ}$ 
in radius centered at the substellar point (solid), longitude
$90^{\circ}$ (dashed), antistellar point (dashed-dotted), and
longitude $-90^{\circ}$ (dotted), all at the equator.   
The profiles demonstrate that dynamics exercises a zeroth-order influence 
on the vertical temperature
structure. On the nightside (dashed-dotted), the intense radiative cooling leads
to a temperature that decreases strongly with height.  
As the equatorial jet transports this air eastward to the terminator 
at longitude $-90^{\circ}$ (dotted), deep regions (0.1 to several bars) have
continued to cool.  Intense stellar heating initiates
as this air reaches the dayside, 
but this has an immediate effect only at low pressures where radiative time constants
are short.  A thermal inversion thus develops at low pressures ($<50\,$mbar).
Westward flow at high latitudes and at pressures $<10\,$mbar also contributes
to this inversion, as such air comes from the hot dayside.  By the time the 
superrotating equatorial jet transports the equatorial
air eastward to the substellar region
(solid), heating has had a substantial effect down to pressures of several
hundred mbar, leading to a deep thermal inversion layer. As the air
reaches the terminator at $90^{\circ}$ longitude, rapid cooling aloft
leads to low temperatures at pressures $<100\,$mbar, but the layer between
0.1--1 bar has longer radiative time constants and, having just crossed
the entire dayside hemisphere, remains warm.  
The key point is that the $T(p)$ structure
varies strongly across the globe and deviates strongly from the predictions 
of 1D radiative-equilibrium models, as previously described in 
\citet{fortney-etal-2006b}.  The day-night temperature
differences are $\sim300\K$ at $1\,$bar, $\sim600\K$ at $100\,$mbar,
and increase to $\sim1000\K$ at the top of the model.

%%%%%%%%%%%%%%%%%
% FIGURE 8
%%%%%%%%%%%%%%%%%
\begin{figure}
\includegraphics[scale=0.63, angle=180]
{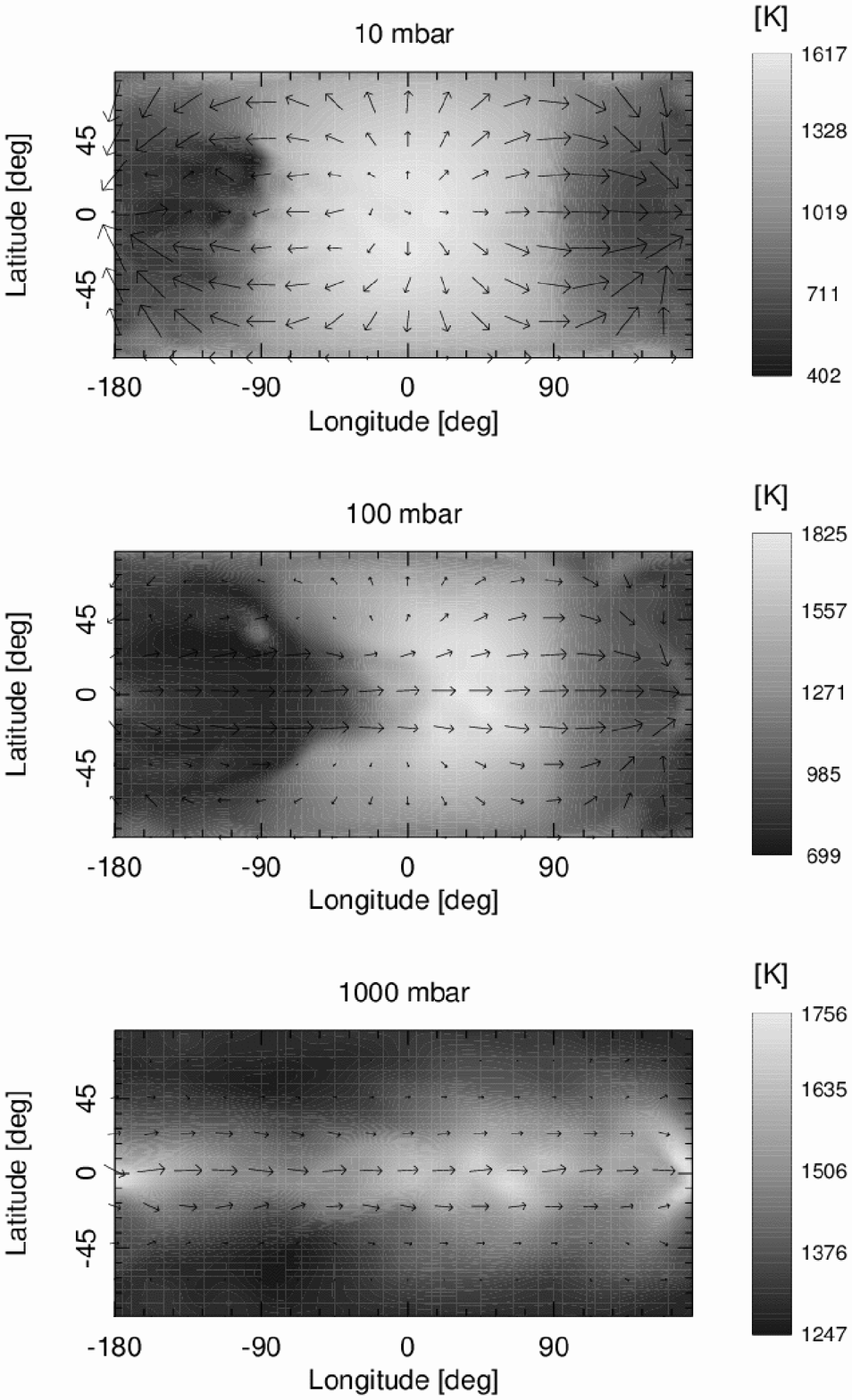}
\caption{Temperature (greyscale) and winds (arrows) for HD209458b 
simulation with half the nominal $\Omega$ at 800 days.
Resolution is $72\times45$.}
\label{209458b-half-omega}
\end{figure}

To test the sensitivity of our results to model resolution, we also performed 
simulations of HD209458b at a lower horizontal resolution of $72\times45$
with 40 layers.  These
simulations produced patterns of wind and temperature that are nearly identical
to Fig.~\ref{209458b-t-winds}.  While we cannot rule out that the 
behavior could change at a resolution higher than considered here, 
this test suggests that, within 
the context of our forcing approach, our horizontal resolutions 
are sufficient to resolve the global-scale flow.  This makes sense 
in light of the fact that (i) the predominant
forcing consists of heating gradients with a large (hemispheric) 
length scale, and (ii) the Rhines length and Rossby deformation 
radius, which determine the predominant
horizontal widths of jets and baroclinic eddies, are comparable 
to the planetary radius.  Our resolutions are sufficient to resolve 
these lengthscales.  This differs from Jupiter, where a relatively 
fine horizontal resolution is required to resolve the small deformation 
radius of $\sim2000\km$.

To determine the influence of planetary rotation rate on the flow geometry, 
we performed HD209458b parameter variations with double and half the 
nominal rotation rate (rotation periods of 1.75 and 7 days, respectively); 
the results are depicted in
Figs.~\ref{209458b-double-omega} and \ref{209458b-half-omega}.  These
models are otherwise essentially identical to the nominal HD209458b case, 
including the use of an insolation pattern relevant for a synchronously 
rotating planet (i.e., the substellar
point is locked at $0^{\circ}$ longitude).  Although the basic flow 
regimes are similar 
to those obtained in the nominal case (Fig.~\ref{209458b-t-winds}), 
a careful comparison reveals several key differences.  Most importantly, 
as can best be seen at $\sim100\,$mbar, 
the superrotating equatorial jet is narrower in the high-rotation-rate case 
(Fig.~\ref{209458b-double-omega}) and wider in the low-rotation-rate case 
(Fig.~\ref{209458b-half-omega}) as compared to the nominal simulation.  The
midlatitude flow is, on balance, weaker at the higher rotation rates.  
Furthermore,
the transition from simple dayside-to-nightside flow to a banded jet pattern
occurs at lower pressures at the higher rotation rates.  
Fig.~\ref{209458b-t-winds}a
and \ref{209458b-double-omega}a develop equatorial flow that is eastward at 
most longitudes; on the other hand, Fig.~\ref{209458b-half-omega}a 
exhibits eastward flow at longitudes 0 to $180^{\circ}$ and westward 
flow at longitudes $-100$ to $0^{\circ}$, leading to a pattern nearly 
symmetric in longitude about the substellar and antistellar
points.  As a result, substantial differences arise in the temperature and
wind patterns at $p\le 100\,$mbar in the nightside quadrant west of 
the antistellar point (longitudes 90--$180^{\circ}$).

%%%%%%%%%%%%%%%%
% FIGURE 9
%%%%%%%%%%%%%%%%
% in emulateapj version I used scale=0.4 without width/height options
% The ``old'' versions (that I cropped from the pdf files) were
% 209458_high_res_omega_2.0x_800d_ubar_crop.ps
% 209458_low_res_omega_0.5x_800d_ubar_crop.ps
\begin{figure}
% This for submitted version:
%\includegraphics[width=5.in, height=2.8in,angle=0]{f9a.eps}
%\includegraphics[width=5.in, height=2.8in, angle=0]{f9b.eps}
%\includegraphics[width=5.in, height=2.8in, angle=0]{f9c.eps}
%
% This for emulateapj version:
\includegraphics[scale=0.4,angle=0]{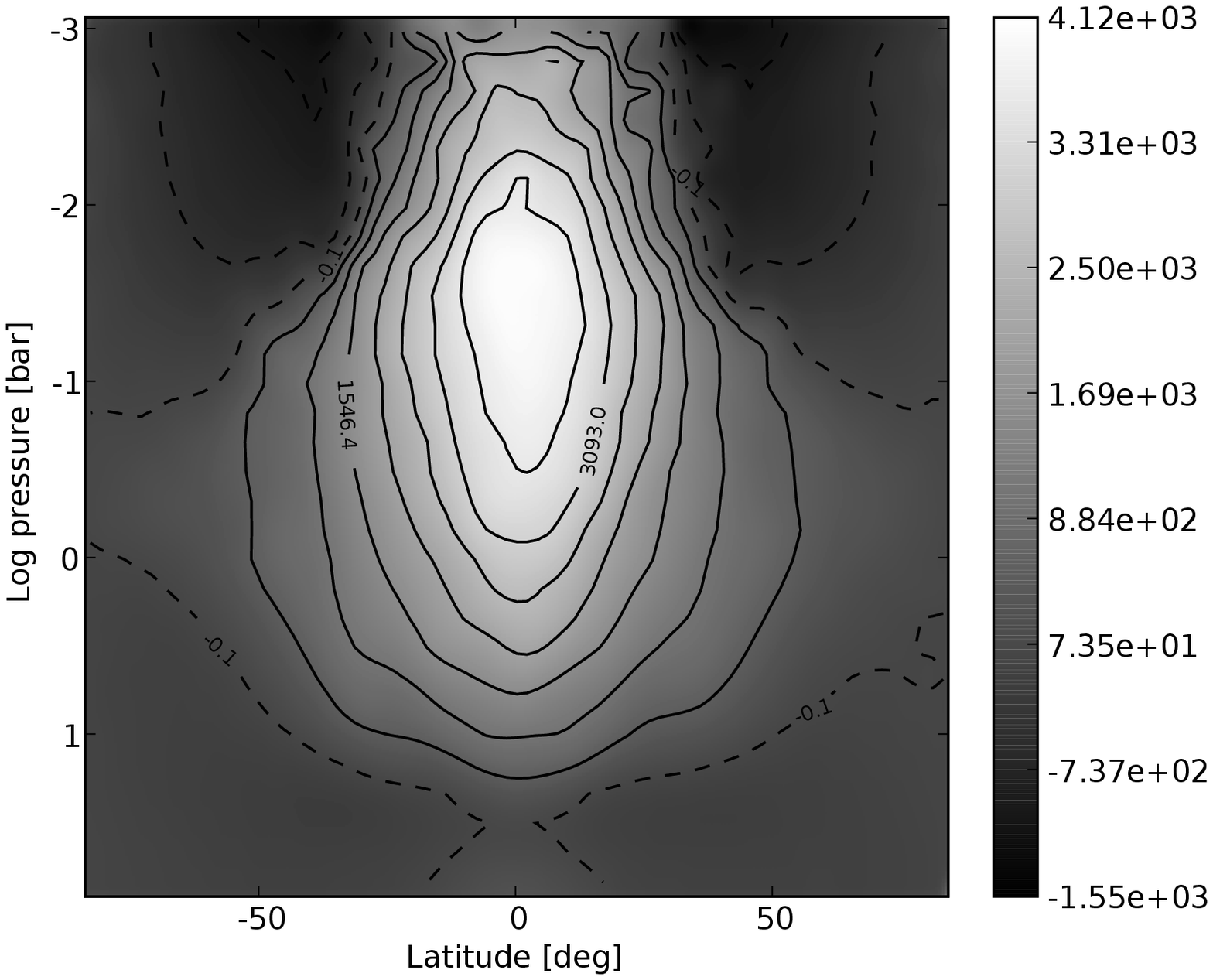}
\includegraphics[scale=0.4, angle=0]{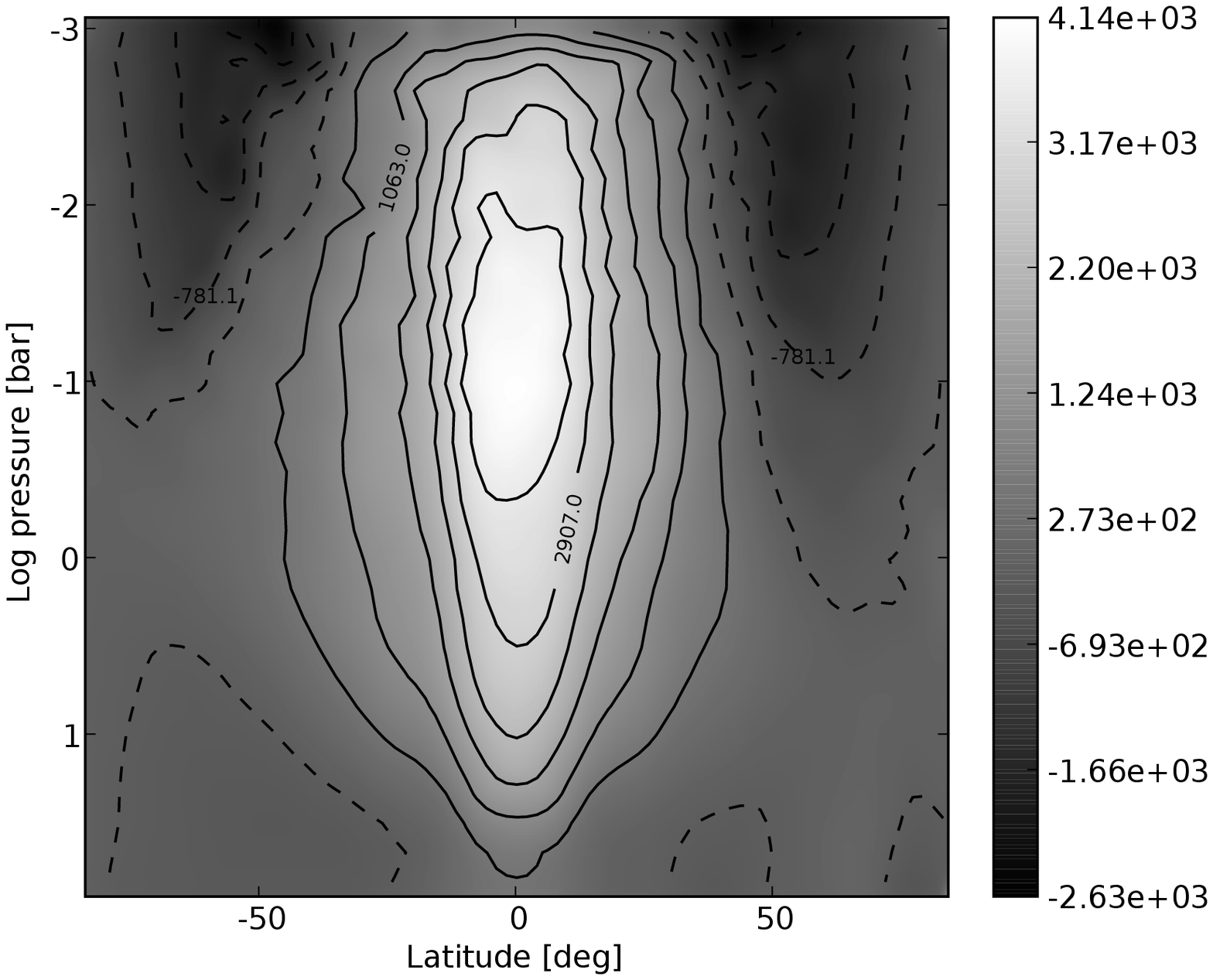}
\includegraphics[scale=0.4, angle=0]{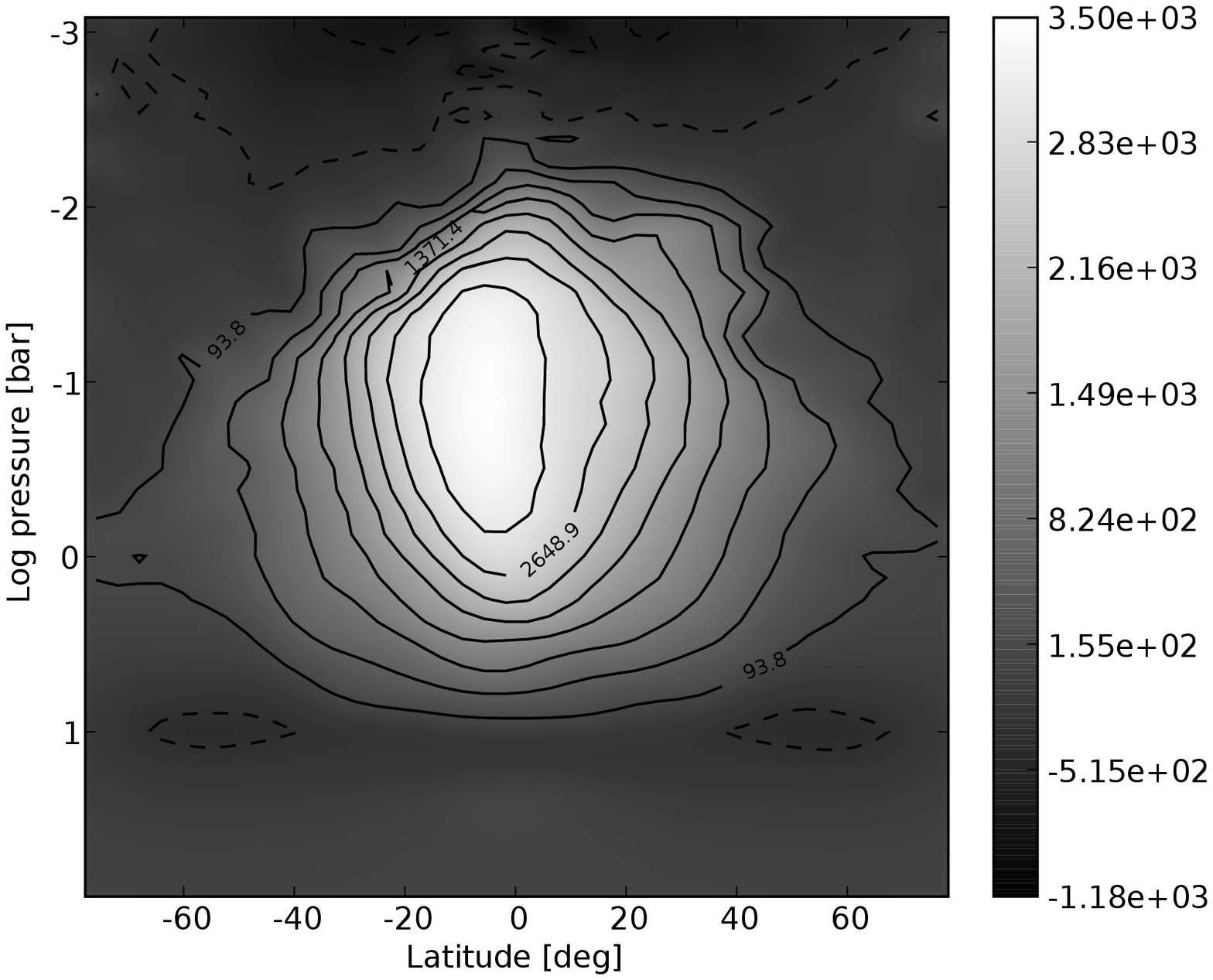}
\caption{
Zonal-mean zonal velocity versus latitude and pressure for HD209458b simulations
at 800 days with nominal rotation rate (top), double the nominal rotation (middle) and
half the nominal rotation (bottom).  Scale bar shows wind speed in $\m\sec^{-1}$;
positive is eastward and negative is westward.  The equatorial jet is wider when the 
rotation rate is lower.
}
\label{ubar}
\end{figure}

These differences in the jet patterns can be better appreciated by
examining the plots of zonally averaged zonal wind shown in Fig.~\ref{ubar}.
The top panel gives the nominal HD209458b simulation 
(as in Fig.~\ref{209458b-t-winds}),
and the middle and bottom panels depict the high- and low-rotation-rate
parameter variations (as in 
Figs.~\ref{209458b-double-omega}--\ref{209458b-half-omega}).
The equatorial jets in the nominal and low-rotation-rate cases have 
latitudinal widths (characterized by full-width-at-half-max) 
approximately 20\% and 40\% greater, respectively, than in the 
high-rotation-rate parameter variation.  This agrees qualitatively with 
the expectation that, for constant wind speed, scale height, and static 
stability, faster rotation implies smaller Rhines length and Rossby 
deformation radius --- and thus should lead to narrower jets.  On the other
hand, the simulated variations in jet width are less than suggested 
by Rhines scaling, which predicts that, at constant wind speed, 
factor-of-four variations in rotation rate should produce 
factor-of-two variations in jet width.  The fact that the forcing 
lengthscale is similar to the jet scale and that it remains constant 
throughout may play a role in muting 
the rotation-rate sensitivity.  Interestingly,
the jets penetrate slightly deeper in the higher-rotation-rate cases.  
The absence of the eastward jet at $p<10\,$mbar in the low-rotation-rate 
case (Fig.~\ref{ubar}c) results from the fact that the flow is largely 
symmetric in longitude about the substellar point
at these low pressures; the eastward and westward branches of the 
circulation cancel out in a zonal average,
leading to weak zonal-mean wind speeds.

We also performed HD209458b simulations with half and double
the nominal gravity (5 and $20\m\sec^{-2}$).  These simulations
are very similar to the nominal case shown in Fig.~\ref{209458b-t-winds},
although the equatorial jet speed in the low-gravity case is
slightly weaker than in the nominal and high-gravity cases.

To determine the sensitivity to the radiative time constant, we performed
an HD209458b simulation analogous to Fig.~\ref{209458b-t-winds} but multiplying
$\tau_{\rm rad}$ by a factor of 10 at all temperatures and pressures. 
This case is particularly relevant because of the uncertainty in our
calculated $\tau_{\rm rad}$. As expected, the winds in this simulation develop more 
slowly and the day-night temperature differences are more muted than in the
nominal cases.  Qualitatively, however, the overall circulation pattern
strongly resembles that in our nominal cases.

%%%%%%%%%%%%%%%
% FIGURE 10
%%%%%%%%%%%%%%
% original file called T_adv-hd189733b-800days.ps
\begin{figure}
\includegraphics[scale=0.45]{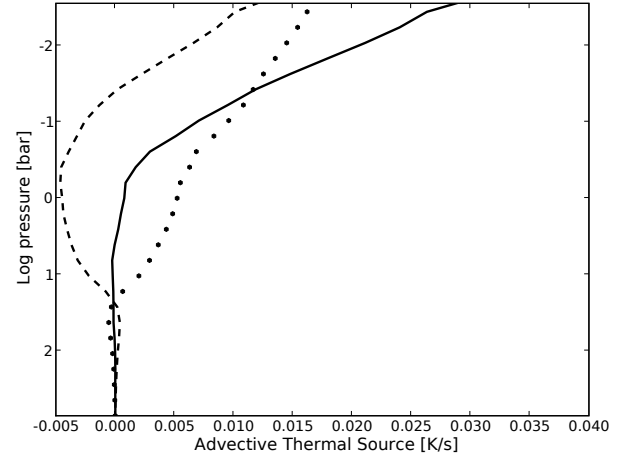}
\caption{Heat source, expressed in $\K\sec^{-1}$, 
associated with day-to-night horizontal
advection across the terminators for our nominal HD189733b simulation. 
Dashed and dotted lines show heat transport
across the dawn and dusk terminators, respectively (third and
fourth terms on the right side of Eq.~\ref{heat-redist}).
Solid line shows their sum.}
\label{fig:heat-redist}
\end{figure}

We wish here to correct the record regarding the issue of
sensitivity to initial conditions. \citet{langton-laughlin-2008}
speculated that the simulated wind speeds reported in 
\citet{cooper-showman-2005} occurred because their simulations were
initialized with the nightside temperature rather than the
global-average temperature.  This is incorrect.  Our
simulation results are insensitive to the initial
temperature profile, particularly at pressures $<10\,$bars; 
cold and hot initial temperatures lead to very similar end states.  
Simulations by \citet{cooper-showman-2006} explicitly demonstrate 
this fact: they initialized their simulations with the global-average
temperature profile rather than the nightside temperature,
yet their simulation results are nearly identical
to those in \citet{cooper-showman-2005}.  This insensitivity occurs 
because the day-night temperature pattern that develops 
in the simulation (which via Eq.~\ref{momentum}--\ref{hydrostatic} 
determines the 
pressure-gradient forces that drive the winds) is dominated
by the radiative heating/cooling patterns and hence quickly
loses memory of its initial condition.  

\subsection{Day-night heat redistribution}

Several recent 1D radiative-transfer models include 
parameterizations of the day-night heat redistribution
by the circulation \citep{burrows-etal-2008, burrows-etal-2006}.
In these studies, an arbitrary fraction of the absorbed 
starlight is removed from the dayside and added to the nightside.
To guide such efforts, we here quantify the magnitude of 
redistribution that occurs in our simulations.  The goal is to
obtain an equation for the mean nightside temperature and its
modification by day-night advection.  Starting with
the thermodynamic energy equation (\ref{energy}), we express
the advection terms in divergence form and invoke the continuity
equation (\ref{continuity}) to cancel out terms, yielding
\begin{equation}
{\partial T\over\partial t} + \nabla\cdot({\bf v}T) 
+ {\partial\over\partial p}(\omega T)  
= {q\over c_p} + {\omega\over\rho c_p}
\label{heat-redist-intermed}
\end{equation}
We then horizontally 
average the equation over the nightside hemisphere, yielding
\begin{equation}
{\partial \langle T\rangle_{\rm night} \over\partial t}=
\left\langle {q\over c_p}\right\rangle_{\rm night}
+ \left\langle {\omega\over \rho c_p} - {\partial\over\partial p}
(\omega T)\right\rangle_{\rm night} - 
\left\langle \nabla\cdot({\bf v}T)\right\rangle_{\rm night}
\end{equation}
where $\langle X\rangle_{\rm night} \equiv 
(2\pi a)^{-1}\int X \, dA$, $a$ is the planetary radius,
the integral is taken over the nightside hemisphere (of area $2\pi a$),
and $dA$ is the increment of horizontal area on the sphere.
Using the divergence theorem, we can then rewrite
$\langle\nabla\cdot({\bf v}T)\rangle_{\rm night}$ 
as the product of the zonal (east-west)
wind and temperature integrated along the day-night terminators:
\begin{equation}
\begin{split}
{\partial \langle T\rangle_{\rm night} \over\partial t}=
\left\langle {q\over c_p}\right\rangle_{\rm night}
+ \left\langle {\omega\over \rho c_p} - {\partial\over\partial p}
(\omega T)\right\rangle_{\rm night} \\
-{1\over2\pi a}\int^{90^{\circ}}_{-90^{\circ}} u(\lambda_{\rm dawn},\phi,p)
T(\lambda_{\rm dawn},\phi,p)\,d\phi \\
+{1\over2\pi a}\int^{90^{\circ}}_{-90^{\circ}} u(\lambda_{\rm dusk},\phi,p)
T(\lambda_{\rm dusk},\phi,p)\,d\phi
\end{split}
\label{heat-redist}
\end{equation}
Here, $\lambda_{\rm dawn}$ and $\lambda_{\rm dusk}$ refer to 
the longitudes of the ``dawn'' and ``dusk'' terminators
($-90^{\circ}$ and $90^{\circ}$), and we have used the fact that
for zero obliquity the terminators lie at constant longitude. 

Equation~\ref{heat-redist} provides a framework for quantifying
the day-night heat redistribution. 
On the right side, the first and second terms in brackets are
the temperature change due to radiative heating/cooling and vertical
motion, respectively.  The final two terms are the ``heat source'' 
associated with thermal-energy transport from dayside to nightside.
\citet{burrows-etal-2008,
burrows-etal-2006} took the approach of introducing an {\it ad hoc}
parameterization for the third and fourth terms on the right side
of Eq.~\ref{heat-redist}.  Here, we explicitly calculate these
terms from our simulations.

Figure~\ref{fig:heat-redist} shows these heat transports versus
pressure for our nominal HD189733b simulation.  The dashed and dotted 
curves show the transport across the terminators at longitude $-90^{\circ}$ 
and $+90^{\circ}$, respectively, and 
the solid curve shows their sum.  Several points deserve mention.
First, the individual transports 
(dashed and dotted curves) are large at low pressure and decay 
toward zero at pressures $>100\,$bars, which results from the 
fact that, in these simulations, 
the circulation is strong aloft and dies out with depth.  

Second, these individual terms (dotted and dashed curves in 
Fig.~\ref{fig:heat-redist}) sum 
constructively at pressures $<50\,$mbar but have opposite signs 
at greater pressures.  This results from the fact that, at low pressure, 
the circulation transports air from dayside to nightside across
{\it both} terminators (see Figs.~\ref{189733b-t-winds}-\ref{209458b-t-winds}), 
leading to a positive heat source for the nightside across both terminators. 
In contrast, at depth, the superrotating equatorial jet transports air from
the dayside to nightside across the dusk terminator (leading to 
a positive contribution in Fig.~\ref{fig:heat-redist}, dotted) but 
transports air from the nightside to dayside across the dawn terminator
(leading to a negative contribution in Fig.~\ref{fig:heat-redist}, dashed).
At the altitudes of the superrotating jet, the day-night heat transports
across the two terminators largely cancel.
As a result of these effects, the {\it net} advective 
heat source is almost zero at pressures exceeding $\sim1\,$bar but
becomes large at lower pressure (solid curve).  The magnitudes 
reach $\sim0.02\K\sec^{-1}$ near the top of the model.

The magnitude of the net advective heat source
(solid curve, Fig.~\ref{fig:heat-redist}) increases monotonically
with altitude.  This differs from the scheme used
by \citet{burrows-etal-2007b, burrows-etal-2008} to predict light curves
and secondary-eclipse depths, which
confined the heat transport to pressures between $\sim0.01$ and 0.1 bars.
Our simulations suggest that a continuous, upward-increasing profile
of temperature modification may be preferable.  

Interestingly, pressures where the advective temperature
modification is greatest (i.e., the top of the model) 
have the largest day-night temperature difference.  
The enormous radiative heating/cooling rates at these levels overwhelm 
the transport, allowing near-radiative-equilibrium conditions (with large 
day-night temperature differences) to be maintained.  In contrast,
at deeper levels, the transport modifies the temperature more weakly,
but the heating/cooling rates are low, allowing relatively homogenized
day-night temperatures to be maintained.   This situation belies 
simple descriptions (common in the literature) of planets with large day-night
temperature differences having ``inefficient'' redistribution and those with
homogenized conditions as having ``efficient'' redistribution.

%%%%%%%%%%%%%%%%%
% FIGURE 11
%%%%%%%%%%%%%%%%%
% originally called fluxes_hd209_800day.eps.
\begin{figure}
\includegraphics[scale=0.50]{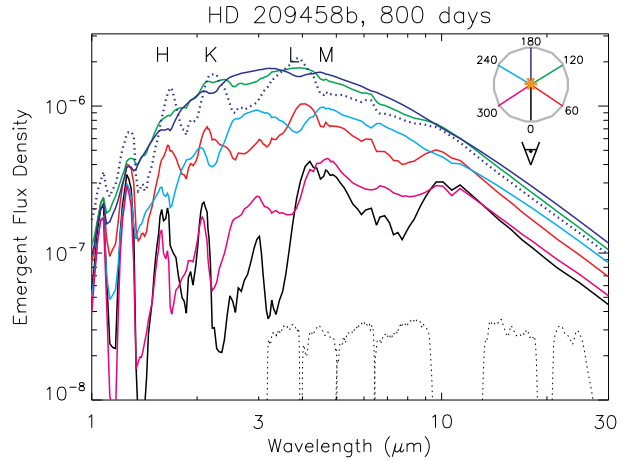}
\caption{Emergent flux density ($\erg \,\sec^{-1}\cm^{-2}\Hz^{-1}$)
from the nominal HD209458b simulation at six phases assuming 
equilibrium chemistry: black --- nightside, as seen during transit; 
red --- $60^{\circ}$ after transit; green --- $120^{\circ}$ after 
transit; dark blue --- dayside, as seen during secondary eclipse;
light blue --- $60^{\circ}$ secondary eclipse; and
magenta --- $120^{\circ}$ after secondary eclipse.
The key in the upper right corner is color-coded with
the spectra to illustrate the sequence.  For reference, the dotted blue 
curve shows the spectrum of a 1D radiative-equilibrium model atmosphere.  
Thin dotted black lines at the bottom of the figure show normalized Spitzer
band passes and the letters at the top show locations of
the $H$, $K$, $L$, and $M$ bands.}
\label{spectrum-209458b}
\end{figure}

%%%%%%%%%%%%%%%%
% FIGURE 12
%%%%%%%%%%%%%%%%
% originally called LCplot_HD189_900d-final.eps.
\begin{figure}
\includegraphics[scale=0.50]{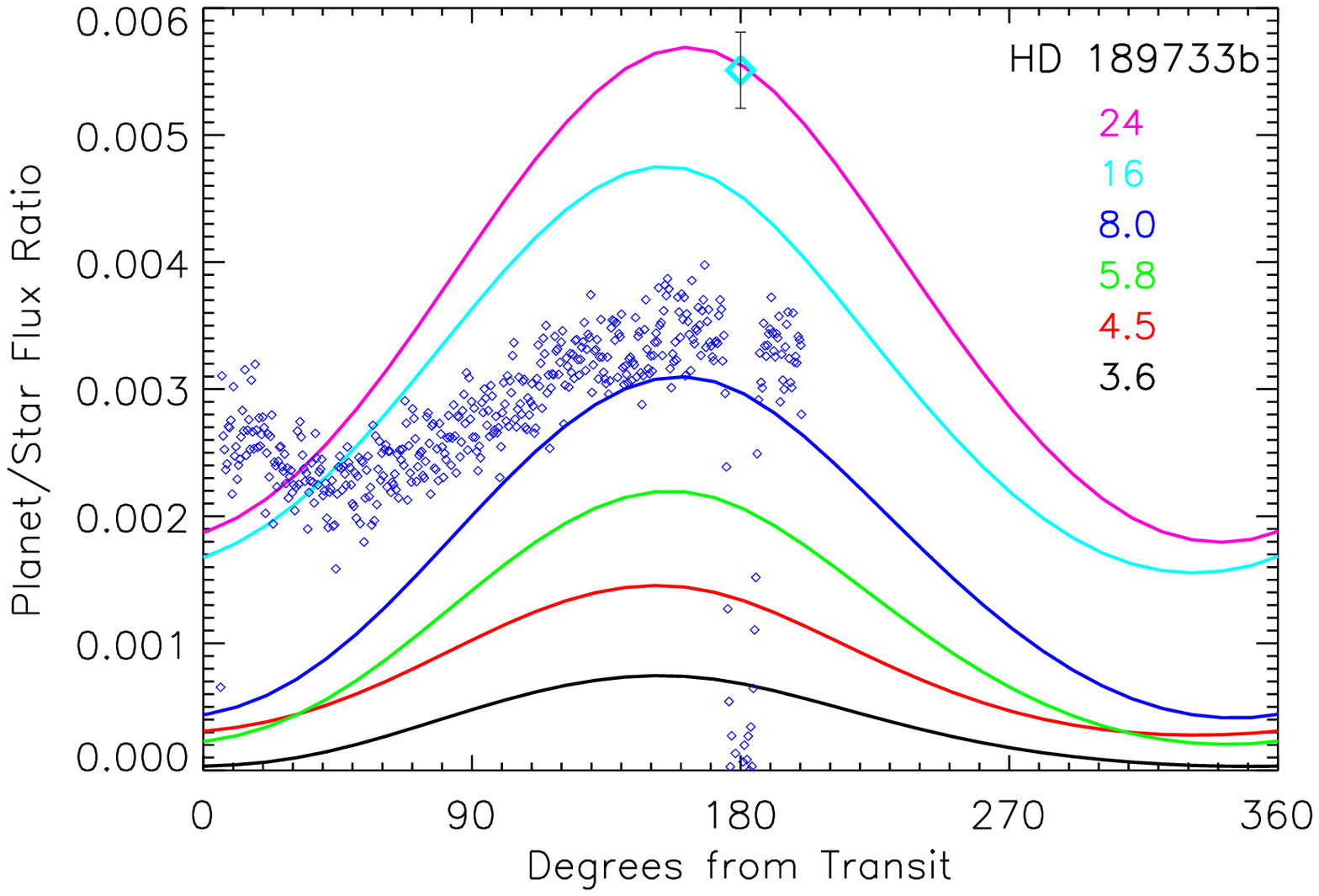}
\caption{Light curves versus orbital phase calculated in Spitzer 
bandpasses for HD189733b.  From top to bottom, the light
curves are for wavelengths of $24\,\mu$m (magenta),
$16\,\mu$m (light blue), $8\,\mu$m (dark blue), $5.8\,\mu$m (green), 
$4.5\,\mu$m (red), and $3.6\,\mu$m (black), respectively.  
Overplotted is the Spitzer 8-$\mu$m light curve
from \citet{knutson-etal-2007b} in dark blue and the 
16-$\mu$m secondary-eclipse depth from \citet{deming-etal-2006} in light blue.
}
\label{lightcurve-189733b}
\end{figure}

%%%%%%%%%%%%%%%%%
% FIGURE 13
%%%%%%%%%%%%%%%%%
% originally called LCplot_HD209_800d-final.eps.
\begin{figure}
\includegraphics[scale=0.50]{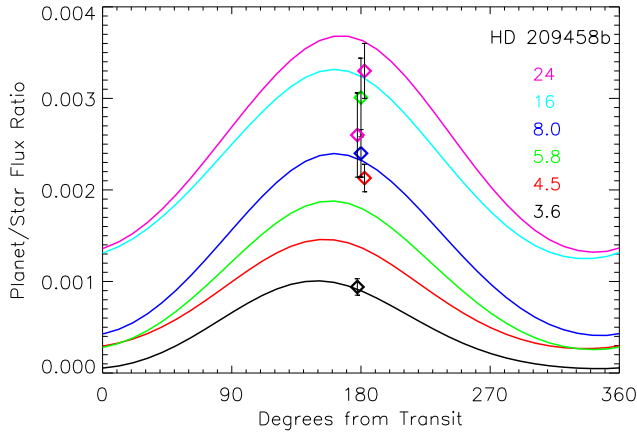}
\caption{Same as in Fig.~\ref{lightcurve-189733b} but for HD 209458b.
Diamonds give measured Spitzer secondary-eclipse depths with 1-$\sigma$ error bars. 
Wavelengths of measurements and light curves are color coded to each other 
and to the numbers given in the plot.  Magenta is $24\,\mu$m
(\citet{deming-etal-2005a} for the lower point and D. Deming, personal
communication, for the upper point), light blue is $16\,\mu$m,
dark blue is $8\,\mu$m \citep{knutson-etal-2007b},
green is $5.8\,\mu$m, red is $4.5\,\mu$m, and black is $3.6\,\mu$m
\citep[all from][]{knutson-etal-2007d}.}
\label{lightcurve-209458b}
\end{figure}

%%%%%%%%%%%%%%%%%%%
% FIGURE 14
%%%%%%%%%%%%%%%%%%%
% originally called day_hd209.eps.
\begin{figure}
\includegraphics[scale=0.50]{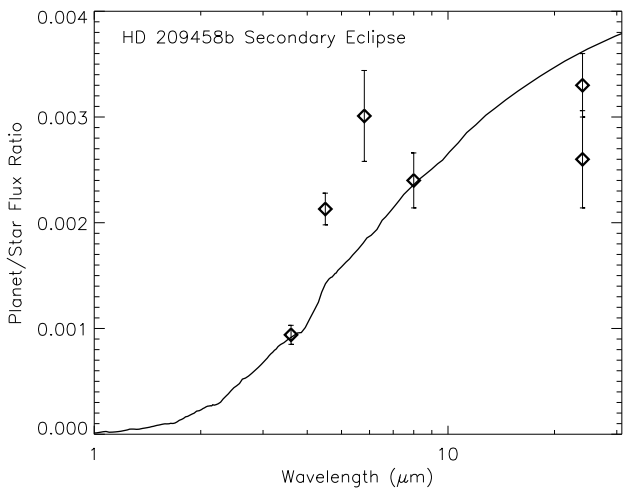}
\caption{Planet-to-star flux ratio versus wavelength for our nominal
HD209458b simulation at the time immediately before/after secondary eclipse.
Points show measured secondary-eclipse depths \citep{deming-etal-2005a,
knutson-etal-2007b, knutson-etal-2007d}. As discussed in the text,
our inability to fit the points at 4.5 and $5.8\,\mu$m probably results
from the lack of a stratosphere in our simulations.}
\label{planet-to-star-209458b}
\end{figure}

\section{Spectra and light curves}

We calculate spectra and light curves from our simulations 
following the methods described in \citet{fortney-etal-2006b}. 
The 3D temperature field from a dynamical simulation
at a given time can be viewed as numerous individual $T(p)$ 
columns, one for each (longitude, latitude) location on the grid.
Our low- and high-resolution simulations contain
3240 and 12,960 such columns, respectively.  At any
orbital phase (except during secondary eclipse), half of these columns will
be visible from Earth.  Assuming solar metallicity, 
we run each such profile through
our radiative-transfer solver (\S 2.2) to determine
the net emergent flux density versus wavelength for that
column. This calculation properly includes the appropriate value
of $\mu_{\oplus}$ for each column, where $\mu_{\oplus}$ is the cosine of the 
angle between local vertical of each column and the line of
sight to Earth; this naturally accounts for any limb 
brightening or darkening.  The emergent flux calculated for each column
is then weighted by the apparent area of that patch of
the planet as viewed from Earth; these are then summed
to obtain the total emergent flux density of the planet
at a given orbital phase.  \citet{fortney-etal-2006b} gives
further details on the method.

Fig.~\ref{spectrum-209458b} depicts
the resulting spectra for our nominal HD209458b case under the 
assumption that local chemical equilibrium holds; qualitatively
similar behavior occurs for HD189733b.  The spectra
vary strongly with orbital phase.  During secondary eclipse
(black solid curves), the emergent flux is low and 
deep absorption bands of CH$_4$ and H$_2$O appear.
Away from transit, as the dayside comes into view, the infrared
flux increases and the absorption bands
become shallower.  Near secondary eclipse (dark blue), the dayside
faces Earth and the spectra exhibit only modest deviations from a blackbody.  
Interestingly, at this phase, the weak spectral features that exist have flipped 
into emission.  

These phase
variations in spectra result directly from the fact that the 
simulated $T(p)$ structure depends strongly on longitude.  The
radiative time constant is short at low pressure and long at 
high pressure (Fig.~\ref{tau-rad}).  Thus, on the nightside,
the air aloft cools more rapidly than the air at depth, 
leading to a temperature that decreases strongly with altitude 
(Fig.~\ref{t-profiles}).  Conversely, on the dayside, the air aloft 
warms more rapidly than the air at depth, leading to a shallower 
structure that is quasi-isothermal or even exhibits a temperature 
inversion where $T$ increases with altitude (Fig.~\ref{t-profiles}).  
This latter behavior is the reason
for the weak emission features in the dayside spectra (dark blue
curves in Fig.~\ref{spectrum-209458b}).

We now turn to infrared light curves. Integrating our
spectra and the \citet{kurucz-1993} model of the star 
HD189733 or HD209458 (as appropriate) over the Spitzer
bandpasses, we calculate the planet-to-star flux ratios
versus orbital phase.  Figures~\ref{lightcurve-189733b}
and \ref{lightcurve-209458b} display these light curves
for equilibrium chemistry for our nominal models of HD189733b and 
H209458b.  As in \citet{fortney-etal-2006b}, the light curves
here show a phase offset that results from the distortion
of the temperature field by dynamics --- most importantly,
the eastward displacement of the hottest region from the
substellar point and coldest region from the antistellar point,
which causes the flux minima and maxima to occur before
transit and secondary eclipse, respectively.
The offset is greatest ($\sim30^{\circ}$) at $3.6\,\mu$m
and smallest ($\sim20^{\circ}$ for HD189733b and
$13^{\circ}$ for HD209458b) at $24\,\mu$m.  As described
in \citet{fortney-etal-2006b}, this wavelength-dependence
results from the pressure-dependence of the photospheres, 
which tend to be at lower pressures for the longer Spitzer 
wavelengths.  In our simulations, lower pressures have
temperature patterns that more closely track the stellar
heating patterns (Figs.~\ref{189733b-t-winds}--\ref{209458b-t-winds}),
leading to the smaller phase shifts.

In our simulations, the infrared light curves reach a steady pattern
relatively quickly.  Light curves calculated at 100 days are
very similar to those calculated at 900 days.

Fig.~\ref{lightcurve-189733b} compares our light curves
to the 8-$\mu$m Spitzer IRAC light curve of HD189733b
\citep{knutson-etal-2007b}.  Interestingly, our simulations
correctly produce the observed offset of the flux maximum from the 
time of secondary eclipse ($16\pm 6^{\circ}$); however, we do not 
explain the location of the flux minimum, which occurs before 
transit in our simulations but after transit in the observations.
To explain this feature would require a cold region to the
{\it west} of the antistellar point.  Because the
low-latitude winds in our simulations are predominantly eastward,
the cold region in our simulation is displaced to the east rather
than the west.  In the context of the simulations presented here, 
simultaneously explaining both an eastward displacement of 
the hot region and a westward displacement of the cold region 
is difficult.  

If the circulation is quasi-steady, one possibility is that 
the dayside and nightside 8-$\mu$m
photospheres sample different pressures; the observed light curve
might be explained if the equatorial jet speed were eastward at the 
pressure of the dayside photosphere and westward at the pressure
of the nightside photosphere.  Another possibility is that a localized
region of strong ascending motion (perhaps driven by absorption or
breaking of atmospheric waves) could transport low-entropy (hence
low-temperature) air to the photosphere level in a localized region, 
perhaps explaining the low-flux region.  Alternatively, perhaps meandering
hot and cold vortices could lead to time-variable temperature
patterns and light curves \citep{cho-etal-2003, rauscher-etal-2008,
rauscher-etal-2007a}; the low-flux region might then result 
from fortuitous placement of a cold vortex to the west of
the antistellar point at the time of the observations.   However,
we caution that the star HD189733 exhibits starspots that could
influence the light curve; it is unclear whether all the 
structure in the light curve results from the planet rather
than variability in the star.  Future observational and
theoretical work may resolve this ambiguity.

Perhaps more importantly, our simulations greatly underpredict
the $8\,\mu$m flux on the planet's nightside 
(Fig.~\ref{lightcurve-189733b}), suggesting that 
our nightside temperatures are too cold at the 8-$\mu$m photosphere.
At the expected 8-$\mu$m photospheric pressure of $\sim30\,$mbar,
our radiative time constants vary from $\sim10^4\sec$
at $1500\K$ to $10^5\sec$ at $500\K$ and rapidly rise to
$10^7\sec$ at temperatures less than $500\K$ (Fig.~\ref{tau-rad}).  
For an equatorial
jet speed of $\sim3\km\sec^{-1}$ relevant for our simulations, 
the time to advect air parcels across a hemisphere is $10^5\sec$;
thus, in our simulations, the nightside temperature at the 8-$\mu$m
photosphere rapidly cools to $\sim500\K$, at which point
further cooling is inhibited by the rapidly increasing 
$\tau_{\rm rad}$.  In contrast, the nightside 8-$\mu$m
brightness temperature inferred from the \citet{knutson-etal-2007b}
light curve is $973\pm33\K$. 

The fact that we overpredict the 8-$\mu$m day-night flux variation
(and underpredict the absolute nightside flux) probably results
from the simplifications inherent to the Newtonian heating/cooling scheme.
Newtonian heating/cooling 
neglects the fact that the radiative equilibrium
temperature and timescale can depend on the atmosphere's dynamical
response, particularly when actual temperatures are far from
radiative equilibrium.  In reality, nonlinearities
not included in the Newtonian relaxation framework can lead to
radiative interactions between vertical levels that are not
accounted for here.  For example, because
of the slow radiative heating rates at deep levels, the temperatures 
at $p\ge1\,$bar become homogenized between day and nightsides;
upwelling infrared radiation from this level will then warm the
entire $p$-$T$ structure at $p < 1\,$bar and lead to nightside
radiative-equilibrium temperatures that --- because of this dynamical 
response --- exceed those in Fig.~\ref{t-eq}.  This could lead
to warmer nightside conditions than predicted here.

Next consider HD209458b (Fig.~\ref{lightcurve-209458b}). From
several brief Spitzer observations at different phases, 
\citet{cowan-etal-2007} place a 2-$\sigma$ upper limit of 0.0015 
on the peak-to-peak flux variation at $8\,\mu$m.  In contrast, 
our light curve calculated under the assumption of equilibrium 
chemistry exhibits a peak-to-peak variation at $8\,\mu$m of 
$\sim0.0020$, modestly higher than the upper limit.  Regarding
secondary-eclipse photometry, we match well the observed eclipse 
depths at 3.5 and $8\,\mu$m \citep{knutson-etal-2007d}.
Two measurements exist of the 24-$\mu$m eclipse depth,
$0.0026\pm00046$ \citep{deming-etal-2005a} and 
$0.0033\pm0.0003$ (D. Deming, personal communication); 
our simulated light curve is marginally consistent with the latter.  On
the other hand, we underpredict the 4.5-$\mu$m and 5.8-$\mu$m
eclipse depths by $\sim40$\% (see also Fig.~\ref{planet-to-star-209458b}).  
The high flux at these
wavelengths has been interpreted as resulting from a 
thermal inversion at pressures $<0.1\,$bar \citep{knutson-etal-2007d,
burrows-etal-2007b}.   Producing this feature may require the 
existence of stratospheric absorbers such as TiO and VO 
\citep{fortney-etal-2007b, burrows-etal-2007b, burrows-etal-2008},
which are not included in the present models.

Although in chemical equilibrium the primary carbon carrier would
be CO on the dayside and CH$_4$ on the nightside, \citet{cooper-showman-2006} 
showed that interconversion between CO and CH$_4$ should become
chemically quenched at low pressure, leading to nearly constant abundances of these
species everywhere above the photosphere.    
For HD209458b, their simulations suggested a quenched CH$_4$ mole 
fraction corresponding to $\sim1$--$2\%$ of the total carbon.  
Although \citet{cooper-showman-2006} did not consider HD189733b specifically, 
their ``cold'' parameter variation has similar temperatures to our HD189733b 
simulation; the quenced CH$_4$ abundance was $\sim20\%$ of the total CO+CH$_4$
in this case.  The specific values should be viewed as quite uncertain.

We thus recalculated the phase-dependent spectra and lightcurves assuming 
constant CO and CH$_4$ mole fractions rather than chemical equilibrium.  
Having only a fraction of the carbon in CH$_4$ on the nightside 
(compared to the equilibrium case where 100\% of the nightside carbon resides in CO) 
lessens the depth of the nightside absorption features at 3.6 and $8\,\mu$m, 
making the nightside brighter at these wavelengths.\footnote{In our
simulations, the dayside fluxes are less sensitive to the CH$_4$/CO ratio 
because of the near-isothermal dayside conditions.}
The effect is modest when the fraction of carbon in methane is 20\% but strong when 
it is only 2\%.  In the latter case, the
day-night flux variation at $3.6\,\mu$m is reduced from a factor of
$\sim10$ to $\sim2$ while that at $8\,\mu$m is reduced from a factor
of $\sim4$ to $\sim2.5$ \citep[see][Fig.~9]{fortney-etal-2006b}.

By lessening the
predicted day-night flux variation, this disequilibrium
effect allows us to fit the upper limit
of 0.0015 on the peak-to-peak planet/star flux variation  
of HD209458b at $8\,\mu$m \citep{cowan-etal-2007}.  It could also potentially help 
provide a better fit to the 8-$\mu$m light curve of HD189733b 
\citep[Fig.~\ref{lightcurve-189733b};][]{knutson-etal-2007b}, 
but only if the planet's methane abundance is relatively small
(perhaps 5\% or less of the total carbon if the metallicity is solar).
Clearly, light curves in the $3.6\,\mu$m Spitzer band (and the 
$4.5\,\mu$m/$3.6\,\mu$m band ratio, which is sensitive to the CH$_4$/CO ratio) 
will provide important constraints on possible disequilibrium chemistry.

\section{Comparison to other work}

\citet{showman-etal-2007} reviewed the various approaches 
adopted for investigating the atmospheric circulation of hot 
Jupiters; here we expand upon key points.

Of the published models, \citet{dobbs-dixon-lin-2008} most 
strongly resembles the approach taken here.  They performed 
3D numerical simulations of the Navier-Stokes equations in the 
low- and mid-latitudes (excluding the poles). 
Radiation was parameterized by flux-limited diffusion using Rosseland-mean 
opacities.   They parameterized stellar irradiation by imposing a
spatially varying temperature at the upper boundary (hot on
the dayside and cold on the nightside); 
downward diffusion of energy at the model top thus induces 
heating on the dayside while upward diffusion out the model top 
induces cooling on the nightside.   Their
nominal simulations adopt opacities relevant to interstellar-medium (ISM)
gas and grains; as they point out, this is probably an overestimate for
planetary atmospheres, where significant particle growth and settling
is expected.  They thus also performed cases with opacities reduced
by factors of 100 and 1000 relative to the ISM and suggest that
100-fold reduction may be most appropriate for hot Jupiters.

For diffusion to be a good approximation, the mean-free path
of the relevant photons must be much less than the length scales
over which the atmospheric properties vary vertically.  This
is an excellent approximation in the deep interior where opacities
are high.  However, the diffusion approximation breaks down near the 
photosphere, where a large 
fraction of infrared photons 
can escape directly to space.  This is the important region
for infrared spectra, light curves, and secondary-eclipse depths.
In this region, radiative fluxes cannot
be represented as a diffusivity times the local temperature 
gradient, as postulated in the diffusion approach; instead, the
radiative transfer involves nonlocal interactions between different
atmospheric levels.   Nevertheless, 
\citet{dobbs-dixon-lin-2008}'s approach, like ours, serves its intended 
purpose of providing an approximate, computationally efficient 
means to force a plausible flow.

A further difference between \citet{dobbs-dixon-lin-2008} and the
present study involves the representation
of dynamics.  They solved the Navier-Stokes equations rather than
the primitive equations; we return to this issue in \S 6.
Additionally, they included the centrifugal
acceleration in the equation of motion; this differs from the
approach taken in most planetary circulation models, which typically
account for the gravitational relaxation of the planetary interior
to the planetary rotation by absorbing the centrifugal force into
the gravity.  
\footnote{For example, a hypothetical rotating planet with no
radiative forcing, no initial horizontal temperature/pressure
gradients (i.e., along the rotationally modified equipotentials),
and zero initial winds should remain motionless over time.  
If the initial condition is spherically symmetric, as assumed by
\citet{dobbs-dixon-lin-2008}, then inclusion of the centrifugal acceleration, 
however, would cause
equatorward motion in such a case --- effectively forcing the
equatorial bulge to develop in the modeled thin atmosphere rather than
throughout the planetary interior.} At the present time, however, 
the greatest source of uncertainty lies in the treatment of radiation
rather than dynamics.

Despite the differences between \citet{dobbs-dixon-lin-2008}'s 
approach and that adopted here, their basic results share strong 
similarities with ours.  Because their case exploring opacities 
100 times less than ISM values 
seems most relevant to hot Jupiters \citep{dobbs-dixon-lin-2008}, 
this is their best simulation to compare with ours.
In agreement with our results, this case develops a banded pattern
at the photosphere with a broad, 
planet-encircling superrotating equatorial jet reaching speeds of 
$3\km\sec^{-1}$ and weaker ($\sim1$--$2\km\sec^{-1}$) westward jets in 
the midlatitudes (see top panels of their Figs.~9--10).  Their day-night 
temperature differences at the photosphere reach $\sim500\K$. 
Although our simulated day-night temperature difference at the 
photosphere differs somewhat from theirs
($\sim800\K$ and $\sim400\K$ for our nominal and $10\times$ nominal
$\tau_{\rm rad}$ cases, respectively), the qualitative similarities between
their simulation and ours are striking given the distinct 
approaches.\footnote{Their cases using ISM opacities exhibit some qualitative
differences in jet structure relative to both their reduced-opacity cases and
our results.   These differences presumably result from the extreme opacity 
in these simulations.}
These similarities, which lend credence to both approaches, are encouraging
because they suggest that the basic flow regime on hot Jupiters
is relatively insensitive to differences in model formulation.

\citet{langton-laughlin-2007} performed numerical simulations of the 
atmospheric flow on hot Jupiters using
the one-layer shallow-water equations, which govern the behavior
of a thin layer of hydrostatically balanced, constant-density fluid
on a sphere.  The momentum and continuity equations are 
\citep[e.g.,][Chapter 3]{pedlosky-1987}
\begin{equation}
{d{\bf v}\over dt}=-g\nabla h - f{\bf k}\times {\bf v}
\end{equation}
\begin{equation}
{\partial h\over\partial t} = -\nabla \cdot(h{\bf v})
\label{sw-continuity}
\end{equation}
where ${\bf v}(\lambda,\phi,t)$ is horizontal velocity, 
$h(\lambda,\phi,t)$ is the thickness of the fluid layer, and $g$
is gravity.  Note that, because of the
constant-density assumption, the temperature is undefined and
there is no thermodynamic-energy equation.   When writing these
equations, \citet{langton-laughlin-2007} replaced $gh$ with $RT$,
causing the continuity equation (Eq.~\ref{sw-continuity}) to 
resemble an energy equation.  This is not a valid procedure, however,
since the shallow-water layer thickness is a representation of mass 
per area between two bounding material surfaces and is thus a distinct 
quantity from temperature;\footnote{Equating $gh$ to $RT$ is equivalent
to assuming $h=RT/g$, which is just the atmospheric pressure scale
height derived for a compressible ideal-gas equation of state.
This is incompatible with the incompressibility assumption on
which shallow-water is based; furthermore, a scale height is defined
as the height between two isobars whereas the shallow-water layer
thickness is the height between two surfaces of constant 
potential-density (e.g., density in the ocean and entropy in the
atmosphere).}   as mentioned above, 
the constant-density assumption prevents temperature from entering 
the shallow-water system.   Their temperature fields are therefore best
interpreted as shallow-water layer thickness. 
\citet{langton-laughlin-2007} forced Eq.~\ref{sw-continuity} using a 
Newtonian-relaxation scheme that adds mass on the dayside and removes 
it on the nightside, which is intended as a simple means to represent 
dayside radiative heating and nightside cooling.  Their simulations
exhibit development of winds reaching $\sim1\km\sec^{-1}$.

In a subsequent study,
\citet{langton-laughlin-2008} performed global, 2D hydrodynamic 
simulations of the flow on hot Jupiters with eccentric orbits.
This system of equations, unlike the shallow-water set, contains independent 
continuity and energy equations; governing variables are ${\bf v}$,
$\rho$, and $T$ over the globe.  They treated radiation using a two-band 
model, one for stellar irradiation and the other for planetary thermal 
radiation under the assumption that the local emission occurs as a blackbody.
Their simulations develop strong lateral temperature 
gradients as one side of the planet is flash heated during periastron
passage; this leads to complex and highly turbulent flow fields as the
resulting vortex structures become dynamically unstable.
This natural development of turbulence from the hemispheric-scale
stellar forcing is an interesting result that has not occurred to date
in models of hot Jupiters with zero eccentricity
\citep{showman-guillot-2002, cooper-showman-2005, cooper-showman-2006,
langton-laughlin-2007, dobbs-dixon-lin-2008}, and it may have
important implications for infrared light curves \citep{langton-laughlin-2008}.
Nevertheless, 2D dynamics differs in important ways from 3D dynamics,
and it remains to be demonstrated whether their adopted 2D model can
reproduce the known flow fields on solar-system planets, which is an
important benchmark for any model.  Exploring
3D circulation models of hot Jupiters on eccentric orbits is clearly
an interesting avenue for future research.

\citet{cho-etal-2003, cho-etal-2008} performed global one-layer simulations
of the flow on hot Jupiters using 
the equivalent-barotropic equations, which are essentially a vertically
averaged version of the 3D primitive equations.  A major difference
with the other studies described here is that \citet{cho-etal-2003, cho-etal-2008}
did not include any radiative heating/cooling; instead, they forced
their flows using a combination of small-scale turbulence in the initial 
condition and a hemispheric-scale pressure deflection intended to 
qualitatively represent the dynamical effects of the day-night heating 
gradient.  In cases when strong initial turbulence was included, their 
flows developed meandering polar vortices, waves, and a high degree of 
turbulent mixing.
The turbulent initial condition that enables these outcomes 
was intended as a parameterization of turbulence generation by 
atmospheric instabilities.  Although all planetary atmospheres are turbulent,
the appropriate turbulent length scales and energetic amplitudes 
remain unknown for hot Jupiters, and may well vary from planet to planet. 
If such turbulence is indeed present at the levels explored by Cho et al., it 
should ideally develop naturally in models that force the flow (for example
with radiative heating/cooling).  So far, however, radiatively forced
investigations of hot Jupiters in circular orbits have exhibited 
relatively steady circulation patterns that lack the degree of
turbulent variability seen in the Cho et al. simulations \citep[][and the
present study]{showman-guillot-2002,
cooper-showman-2005, langton-laughlin-2007, dobbs-dixon-lin-2008},
with the notable exception of \citet{langton-laughlin-2008}'s 2D 
study of hot Jupiters in highly eccentric orbits. 
It will be interesting to see if this continues to be the case as 
a wider range of hot Jupiters is explored and model
resolutions and forcing realism improve over time.

In contrast to the above dynamical models, \citet{iro-etal-2005} and 
\citet{burrows-etal-2006, burrows-etal-2007b, burrows-etal-2008}
explored models of hot Jupiters
that used realistic radiative-transfer schemes but made {\it ad hoc} assumptions
for the effect of dynamics on the thermal budget. \citet{iro-etal-2005}
assumed atmospheric winds that correspond to solid-body rotation;
the speed at the equator was a free parameter varied from 
0.5--$2\km\sec^{-1}$.  \citet{burrows-etal-2006, burrows-etal-2007b, 
burrows-etal-2008} simply removed an arbitrary fraction of the 
stellar irradiation from the dayside and placed it on the nightside
to mimic the effect of atmospheric winds.  This is a novel
approach that allows the exploration of how (for example) day-night
temperature differences should depend on the day-night heat transport.
As these authors are careful to point out, however, it is not
a rigorous treatment of dynamics.

\section{The validity of hydrostatic balance}

A scaling analysis demonstrates that local hydrostatic balance is approximately
valid for the large-scale flow on hot Jupiters.  It is important to emphasize
that the local-hydrostatic-balance assumption in the primitive 
equations derives from the
assumption of large aspect ratio and {\it not} from
any assumption on wind speed.  However, the fact that estimated
wind speeds on hot Jupiters are several ${\rm km}\,{\rm sec}^{-1}$, which
is close to the 3-km$\,{\rm sec}^{-1}$ speed of sound in these
atmospheres, suggests that we consider the validity of the
primitive equations in the hot-Jupiter context. The full Navier-Stokes
vertical momentum equation can be written
\begin{equation}
{\partial w\over\partial t} + {\bf v}\cdot\nabla w =-{1\over\rho}
{\partial p\over\partial z} - g + 2 u \Omega \cos\phi
\label{vert-mom}
\end{equation}
where $w$ is vertical wind speed, $t$ is time, ${\bf v}$ is the
horizontal wind velocity, and $u$ is the east-west wind speed.  
The background static hydrostatic balance is irrelevant
to atmospheric circulation and can be removed from the equation.  Define
 $p=p_0(z)+p'$ and $\rho=\rho_0(z)+\rho'$ where $p_0$ and $\rho_0$
are the time-independent basic-state pressure and density and, 
by construction, ${\partial p_0/\partial z}\equiv -\rho_0 g$.  
Primed quantities
are the deviations from this basic state caused by dynamics.
Substituting these expressions into Eq.~\ref{vert-mom}, we can
rewrite the equation as
\begin{equation}
{\partial w\over\partial t} + {\bf v}\cdot\nabla w =-{1\over\rho}
\left({\partial p'\over\partial z} - \rho'g\right) + 2 u \Omega \cos\phi
\label{vert-mom2}
\end{equation}
where the basic-state hydrostatic balanced has been subtracted off.
The terms in parentheses give only the flow-induced contributions 
to vertical pressure gradient and weight 
(they go to zero in a static atmosphere).

For local hydrostatic balance to be a reasonable approximation, the
terms in parenthesis must be much larger than the other terms
in the equation.  The magnitude of $\partial w/\partial t$ is 
approximately $w/\tau$, where $\tau$ is a flow evolution timescale, 
and the magnitude of ${\bf v}\cdot\nabla w$ is the greater of
$uw/L$ and $w^2/H$ where $L$ is the horizontal flow scale and
$H$ is the vertical flow scale.  For global-scale flows,
$L\sim10^7$--$10^8\,$m and $\tau\sim10^4$--$10^5\,$sec.  
The large-scale
flow varies vertically over a scale height $H\sim300\,{\rm km}$ 
\citep{showman-guillot-2002, cooper-showman-2005, 
cooper-showman-2006, dobbs-dixon-lin-2008}.  
For the simulated flow regime in \citet{cooper-showman-2005, cooper-showman-2006} 
and \citet{dobbs-dixon-lin-2008},
${\bf v} \sim u \sim 3\,{\rm km}\,{\rm sec}^{-1}$, $g\sim10\,{\rm m}
\,{\rm sec}^{-2}$, $\Omega\sim 2\times10^{-5}\,{\rm sec}^{-1}$, 
and $w\sim10$--$100\,{\rm m}\,{\rm sec}^{-1}$.
With these values, we find that $\partial w/\partial t
\le 10^{-2}\,{\rm m}\,{\rm sec}^{-2}$, $uw/L\le0.03\,{\rm m}\,{\rm sec}^{-2}$,
$w^2/H\le0.03\,{\rm m}\,{\rm sec}^{-2}$, and 
$\Omega u \sim 0.1\,{\rm m}\,{\rm sec}^{-2}$.
In comparison, for a hot Jupiter with day-night temperature
differences of several hundred K, the flow-induced hydrostatic terms 
$\rho'g/\rho$ and $\rho^{-1}\partial p'/\partial z$ are 
each $\sim10\,{\rm m}\,{\rm sec}^{-2}$.

This analysis implies that, for global-scale hot-Jupiter flows
at the atmospheric pressures considered in our model ($p>1\,$mbar), 
the greatest departure from hydrostaticity results from the 
vertical Coriolis force,
which causes a $\sim1$\% deviation from hydrostatic balance.
The acceleration terms on the left side of Eq.~\ref{vert-mom2}
(which are necessary for vertically propagating sound
waves) cause a $\sim0.3\%$ deviation from hydrostatic balance.
Hydrostatic balance is thus a reasonable approximation for
the large-scale flow.  Nonhydrostatic effects of course become
important at small scales, and it is conceivable that these effects
interact with the large-scale flow in nontrivial ways.
For hot Jupiters, the acceleration terms on the left side
of Eq.~\ref{vert-mom2} only become important for
structures with vertical or horizontal scales less than
$\sim30\,$km and $\sim500$--$1000\,$km, respectively.  
In a numerical model that solved the
full Navier-Stokes equations, the grid resolution
would have to be substantially finer than these values for
the nonhydrostatic behavior to be accurately 
represented.  A full Navier-Stokes solution with a
coarse resolution would effectively be resolving just
the global-scale hydrostatic component of the flow.

Despite the above, some hot Jupiters may be losing mass from the top
of their atmospheres \citep{vidal-madjar-etal-2003, vidal-madjar-etal-2004},
which suggests that hydrostatic equilibrium should break down
at extremely low pressures (above the top of our model)
where this outflow occurs.

We wish here to clarify the issue of vertical velocities.  
\citet{dobbs-dixon-lin-2008} suggested that vertical motions
are faster in the full Navier-Stokes system than in the primitive
equations because of the different vertical momentum equations
in these systems.  Certainly, velocities in a convection zone
require use of a momentum equation that includes vertical 
accelerations.  However, in statically stable atmospheres, the
primary control over the vertical velocity in
a quasi-steady overturning circulation (relevant to the regime simulated 
here) is {\it not} the vertical momentum equation but the thermodynamic
energy equation (Eq.~\ref{energy}).  The reason is that, in this
situation, the radiative heating/cooling rate controls the speed of 
vertical motion.  Because entropy increases with height in a statically
stable atmosphere, adiabatic expansion/contraction in ascending (descending)
air causes temperature at a given height to decrease (increase) over time.
In the absence of radiation, such steady flow patterns are unsustainable 
because they induce density perturbations that resist the motion (i.e.
ascending air becomes denser and descending air becomes less dense than
the surroundings at that altitude).
Thus, steady vertical motion in a stable atmosphere can only occur
as fast as radiation can remove the temperature perturbations
caused by the adiabatic ascent/descent.  This provides a fundamental
constraint on vertical motion that applies equally to the primitive
and the Navier-Stokes equations.

To quantify this idea, consider Eq.~\ref{energy}, which can
be written
\begin{equation}
{\partial T\over\partial t} + {\bf v}\cdot \nabla T - \omega{H^2 N^2
\over R p} = {q\over c_p}
\label{energy2}
\end{equation}
where $H$ is the vertical scale height and $N$ is the Brunt-Vaisala
frequency (the oscillation frequency for a vertically displaced
air parcel in a statically stable atmosphere). An upper limit on the 
attainable vertical velocity in a quasi-steady circulation in
a statically stable atmosphere results 
from equating the right side to the third term on the left side.
One then obtains a peak vertical velocity (measured
in $\Pa\sec^{-1}$)
\begin{equation}
\omega \sim {q\over c_p}{R p \over H^2 N^2}
\label{vert-velocity1}
\end{equation}
which can be approximately expressed as a peak vertical velocity 
in $\m\sec^{-1}$
\begin{equation}
w\sim {q\over c_p}{R\over H N^2}
\label{vert-velocity2}
\end{equation}
 The equation of course couples to the rest of the 
dynamics via $N^2$ and $q/c_p$. For hot Jupiter photospheres, 
where $q/c_p\sim10^{-2}\K\sec^{-1}$, $H\sim300\km$, and 
$N\sim0.001$--$0.01\sec^{-1}$ depending on temperature, gravity, 
and thermal gradient, we then expect $w\sim10$--$100\m\sec^{-1}$ 
or less \citep{showman-guillot-2002}. 
Importantly, Eq.~\ref{vert-velocity2} makes no assumption 
about the form of the vertical momentum equation (hydrostatic versus 
Navier-Stokes); only quasi-steady flow and a statically stable atmosphere 
were invoked. 
Within the context of the static stabilities and heating rates assumed above, 
these vertical velocities should therefore apply to statically stable 
atmospheres regardless of whether they are modeled with the primitive or
Navier-Stokes equations.  On the other hand, because of the intense nightside
cooling, some hot Jupiters may develop convection zones near the photosphere 
on the nightside.  The primitive equations cannot capture the small-scale
convective vertical velocities in this situation. Nevertheless, our simulations
of HD209458b and HD189733b appear not to predict such detached convection regions.

\label{section:hydrostatic}

\section{Conclusions}

We presented three-dimensional numerical simulations of
the atmospheric circulation on HD209458b and HD189733b; 
calculated infrared spectra and light curves from the simulated
temperature patterns; and compared 
these observables with available measurements.  The simulations
were forced with a simplified Newtonian heating/cooling
scheme.  Within the context of this approach, we have
improved on earlier work by calculating for use in this
scheme the true radiative-equilibrium temperatures as a function of
longitude, latitude, and pressure;  likewise, we calculated
the radiative time constants as a function of both temperature
and pressure over the 3D grid.   These radiative time constants
are generally consistent with earlier estimates \citep{iro-etal-2005}
but span a much greater range of pressure and especially temperature.

In our simulations, HD209458b
and HD189733b develop similar circulation patterns, 
although the latter has a global-average temperature 
$\sim200$--$300\K$ cooler than the former (a result of the 
lower stellar flux).
Consistent with earlier work \citep{cooper-showman-2005,
cooper-showman-2006}, our simulations show that,
at low pressure ($<10\,$mbar), the circulation transports air from 
dayside to nightside, both along the equator and over the poles.
At high pressure ($>100\,$mbar), a banded structure emerges
with a broad, fast (3--$4\km\sec^{-1}$) eastward equatorial jet
flanked by weaker westward flow at high latitudes.  The
day-night temperature difference varies strongly with height
and reaches 500--$1000\K$ above the photosphere.  Near the
photosphere, the dynamics distorts the temperature pattern
in a complex manner; consistent with our earlier work, this
generally includes an eastward displacement of the hottest
regions from the substellar point.  Importantly, dynamics
also pushes the vertical temperature profile $T(p)$
far from radiative-equilibrium.  The temperature decreases 
strongly with altitude on the nightside but becomes quasi-isothermal, 
or even exhibits an inversion layer, on the dayside.

Our calculated spectra, calculated assuming a cloud-free
atmosphere with solar metallicity (appropriately modified
for rainout) show that the nightside should exhibit deep
absorption bands.  On the dayside, however, the deviations 
of the spectrum from a blackbody become modest due to the 
quasi-isothermal structure.  Nevertheless, in some simulations 
this dayside structure contains weak {\it emission} features, which 
result from the existence of a thermal inversion in these simulations.

Our light curves calculated in Spitzer bandpasses show that,
for HD189733b, 
we correctly explain the phase offset of the 
flux peak that occurs before the secondary
eclipse measured in the 8-$\mu$m light curve of 
\citet{knutson-etal-2007b}, but we fail to explain the flux minimum 
that occurs after transit and we overpredict the total flux
variation.  For HD209458b, we match the Spitzer IRAC
3.6 and 8-$\mu$m secondary-eclipse depths, and marginally
fit the latter of two $24\,\mu$m MIPS measurements, but we 
underpredict the 4.5 and $5.8\,\mu$m IRAC secondary-eclipse
depths.  This
probably results from the lack of a strong stratosphere
that seems to be implied by these observations.  Future
simulations that include a realistic representation of 
radiative transfer, including the possibility of TiO
and VO opacity for the case of HD209458b, may be needed
to better fit these diverse observations.

Finally, we presented a scaling analysis suggesting that
the large-scale ($\ge10^7\m$) flow on the hot Jupiters explored 
here should be close to local hydrostatic balance; on these large
scales, deviations from hydrostaticity are small.   This result
lends support to the primitive-equation approach adopted
here.  At the present time, the primary source of uncertainty
is not the dynamical scheme but the representation of the
forcing that drives the flow.  

So far, all published studies of the circulation on hot Jupiters 
have forced the flow using relatively severe approximations to 
the radiative transfer or excluded heating/cooling entirely.
Inclusion of realistic radiative transfer is an important goal
for future work.  Other areas for improvement include considering
cloud/haze formation, alternate compositions (supersolar metallicity
and the possibility of TiO and VO), and exploring a wider range
of planetary parameters to capture the diversity of the growing
number of known hot Jupiters.

%%%%%%%%%%%%%%%%%%%%%
% End document body %
%%%%%%%%%%%%%%%%%%%%%

% Acknoledgements section
\acknowledgements
This research was supported by NASA Planetary Atmospheres
grants NNX07AF35G and NNG06GF28G and NASA GSRP NGT5-50462 to APS.   
CSS was supported in part by an appointment to the NASA
Postdoctoral Program at the Astrobiology Institute at the 
University of Arizona, administered by Oak Ridge Associated
Universities through a contract with NASA.

\bibliographystyle{apj}
%\bibliography{showman-bib}

%%%%%%%%%%%%%%%%%%%%%%%%%%%% TABLES %%%%%%%%%%%%%%%%%%%%%%%%%%%%%%%%%%%%%

\begin{deluxetable}{ccccccccccc}
\tabletypesize{\footnotesize}   % for emulateapj version
%\tabletypesize{\tiny} % for submitted version
%\rotate
%\center
\tablecolumns{12}
\tablewidth{0pc}
\tablecaption{Radiative Time Constants for HD 209458b}
\tablehead{
\colhead{P (bar)} & \colhead{400} & \colhead{600} & \colhead{800} & \colhead{1000} & \colhead{1200} & \colhead{1400} & \colhead{1600} & \colhead{1800} & \colhead{2000} & \colhead{2200}  }
\startdata

   0.00066 & 6.05e+05 & 1.16e+04 & 6.36e+03 & 4.85e+03 & 2.33e+03 & 1.19e+03 & 7.38e+02 & 5.40e+02 & 4.38e+02 & 3.83e+02\\
   0.00110 & 6.16e+05 & 1.51e+04 & 8.24e+03 & 6.50e+03 & 2.99e+03 & 1.56e+03 & 9.47e+02 & 6.60e+02 & 5.04e+02 & 4.11e+02\\
   0.00182 & 5.99e+05 & 1.99e+04 & 1.03e+04 & 8.10e+03 & 3.90e+03 & 2.03e+03 & 1.24e+03 & 8.52e+02 & 6.47e+02 & 5.25e+02\\
   0.00302 & 5.20e+05 & 2.69e+04 & 1.30e+04 & 1.05e+04 & 5.12e+03 & 2.76e+03 & 1.68e+03 & 1.15e+03 & 8.72e+02 & 7.11e+02\\
   0.00501 & 4.49e+05 & 3.51e+04 & 1.56e+04 & 1.21e+04 & 6.36e+03 & 3.49e+03 & 2.14e+03 & 1.52e+03 & 1.21e+03 & 1.04e+03\\
   0.00832 & 4.20e+05 & 4.75e+04 & 1.97e+04 & 1.50e+04 & 8.58e+03 & 4.80e+03 & 2.89e+03 & 2.12e+03 & 1.85e+03 & 1.82e+03\\
   0.01380 & 4.21e+05 & 6.34e+04 & 2.59e+04 & 1.78e+04 & 1.07e+04 & 6.39e+03 & 3.96e+03 & 2.85e+03 & 2.59e+03 & 2.72e+03\\
   0.02291 & 4.57e+05 & 8.48e+04 & 3.37e+04 & 2.29e+04 & 1.52e+04 & 9.13e+03 & 5.57e+03 & 3.93e+03 & 3.49e+03 & 3.66e+03\\
   0.03802 & 5.49e+05 & 1.16e+05 & 4.72e+04 & 2.84e+04 & 1.82e+04 & 1.26e+04 & 7.14e+03 & 5.39e+03 & 4.79e+03 & 5.06e+03\\
   0.06310 & 6.72e+05 & 2.45e+05 & 8.12e+04 & 3.52e+04 & 2.49e+04 & 1.77e+04 & 1.03e+04 & 7.87e+03 & 6.97e+03 & 7.42e+03\\
   0.10471 & 7.98e+05 & 4.16e+05 & 1.45e+05 & 5.89e+04 & 3.45e+04 & 2.61e+04 & 1.61e+04 & 1.17e+04 & 1.01e+04 & 1.09e+04\\
   0.17378 & 8.85e+05 & 5.81e+05 & 2.76e+05 & 1.33e+05 & 5.60e+04 & 4.03e+04 & 2.69e+04 & 1.92e+04 & 1.69e+04 & 1.77e+04\\
   0.28840 & 9.35e+05 & 7.01e+05 & 4.67e+05 & 2.51e+05 & 1.04e+05 & 6.93e+04 & 4.76e+04 & 3.52e+04 & 2.67e+04 & 2.89e+04\\
   0.47863 & 9.14e+05 & 7.71e+05 & 6.28e+05 & 4.84e+05 & 2.40e+05 & 1.30e+05 & 9.06e+04 & 6.68e+04 & 4.92e+04 & 4.82e+04\\
   0.79433 & - & - & 1.02e+06 & 9.15e+05 & 6.17e+05 & 2.91e+05 & 1.82e+05 & 1.32e+05 & 1.03e+05 & 1.28e+05\\
   1.31826 & - & - & - & 1.63e+06 & 1.50e+06 & 6.82e+05 & 4.03e+05 & 3.00e+05 & 2.43e+05 & 5.90e+05\\
   2.18776 & - & - & - & 3.02e+06 & 2.99e+06 & 1.76e+06 & 1.00e+06 & 7.67e+05 & 6.93e+05 & 7.37e+05\\
   3.63078 & - & - & - & - & 5.79e+06 & 4.59e+06 & 2.61e+06 & 2.06e+06 & 1.70e+06 & 1.66e+06\\
   6.02560 & - & - & - & - & - & 8.87e+06 & 8.06e+06 & 6.51e+06 & 6.19e+06 & 8.65e+06\\
  10.00000 & - & - & - & - & - & - & - & 2.21e+07 & 2.17e+07 & 3.60e+07\\
  16.59590 & - & - & - & - & - & - & - & - & 7.01e+07 & 9.39e+07\\

\enddata
\tablecomments{Column headers are temperatures in Kelvin.  Values are time constants in seconds.  Note that no values were calculated in the lower left corner here and in Table 2 because the radiative-equilibrium $p$-$T$ profiles used
to calculate $\tau_{\rm rad}$ (Fig.~\ref{t-eq}) do not access these 
pressure/temperature combinations.}
\label{tau209}
\end{deluxetable}

\begin{deluxetable}{cccccccccc}
\tabletypesize{\footnotesize}   % for emulateapj version
%\tabletypesize{\tiny}          % for submitted version
%\rotate
%\center
\tablecolumns{11}
\tablewidth{0pc}
\tablecaption{Radiative Time Constants for HD 189733b}
\tablehead{
\colhead{P (bar)} & \colhead{400} & \colhead{600} & \colhead{800} & \colhead{1000} & \colhead{1200} & \colhead{1400} & \colhead{1600} & \colhead{1800} & \colhead{2000}  }
\startdata

   0.00066 & 2.58e+06 & 7.40e+03 & 4.84e+03 & 3.33e+03 & 1.61e+03 & 7.84e+02 & 4.25e+02 & 2.45e+02 & 1.49e+02\\
   0.00110 & 1.87e+06 & 9.32e+03 & 6.48e+03 & 4.25e+03 & 2.08e+03 & 1.03e+03 & 5.73e+02 & 3.38e+02 & 2.09e+02\\
   0.00182 & 1.53e+06 & 1.19e+04 & 8.42e+03 & 5.44e+03 & 2.65e+03 & 1.34e+03 & 7.58e+02 & 4.58e+02 & 2.91e+02\\
   0.00302 & 1.11e+06 & 1.54e+04 & 1.07e+04 & 7.13e+03 & 3.43e+03 & 1.78e+03 & 1.05e+03 & 6.66e+02 & 4.48e+02\\
   0.00501 & 9.54e+05 & 1.94e+04 & 1.22e+04 & 8.34e+03 & 4.26e+03 & 2.28e+03 & 1.35e+03 & 8.60e+02 & 5.76e+02\\
   0.00832 & 1.05e+06 & 2.54e+04 & 1.42e+04 & 1.01e+04 & 5.50e+03 & 3.00e+03 & 1.77e+03 & 1.11e+03 & 7.24e+02\\
   0.01380 & 1.34e+06 & 3.27e+04 & 1.69e+04 & 1.14e+04 & 6.79e+03 & 3.82e+03 & 2.25e+03 & 1.39e+03 & 8.95e+02\\
   0.02291 & 1.78e+06 & 4.30e+04 & 2.05e+04 & 1.27e+04 & 8.95e+03 & 5.31e+03 & 2.93e+03 & 1.67e+03 & 9.79e+02\\
   0.03802 & 1.86e+06 & 5.77e+04 & 2.56e+04 & 1.44e+04 & 1.12e+04 & 7.00e+03 & 3.94e+03 & 2.18e+03 & 1.20e+03\\
   0.06310 & 2.22e+06 & 8.71e+04 & 3.64e+04 & 2.08e+04 & 1.39e+04 & 8.90e+03 & 5.38e+03 & 3.12e+03 & 1.80e+03\\
   0.10471 & 2.55e+06 & 1.44e+05 & 5.11e+04 & 3.02e+04 & 1.76e+04 & 1.33e+04 & 7.97e+03 & 5.15e+03 & 3.39e+03\\
   0.17378 & 3.55e+06 & 2.43e+05 & 1.03e+05 & 5.55e+04 & 2.74e+04 & 1.88e+04 & 1.23e+04 & 8.52e+03 & 6.20e+03\\
   0.28840 & 4.55e+06 & 7.25e+05 & 2.40e+05 & 1.00e+05 & 5.43e+04 & 3.01e+04 & 1.97e+04 & 1.42e+04 & 1.07e+04\\
   0.47863 & 5.60e+06 & 2.17e+06 & 4.30e+05 & 2.33e+05 & 9.63e+04 & 4.94e+04 & 3.52e+04 & 2.58e+04 & 2.02e+04\\
   0.79433 & 6.64e+06 & 3.78e+06 & 9.20e+05 & 5.28e+05 & 2.15e+05 & 1.03e+05 & 6.57e+04 & 4.76e+04 & 3.63e+04\\
   1.31826 & 7.61e+06 & 5.50e+06 & 3.39e+06 & 1.47e+06 & 5.06e+05 & 2.58e+05 & 1.30e+05 & 9.42e+04 & 7.01e+04\\
   2.18776 & 9.59e+06 & 7.73e+06 & 5.87e+06 & 4.02e+06 & 1.34e+06 & 6.56e+05 & 2.89e+05 & 2.07e+05 & 1.52e+05\\
   3.63078 & 2.52e+07 & 2.03e+07 & 1.47e+07 & 9.04e+06 & 3.74e+06 & 1.68e+06 & 6.95e+05 & 5.34e+05 & 4.36e+05\\
   6.02560 & 5.96e+07 & 5.22e+07 & 3.82e+07 & 2.42e+07 & 9.70e+06 & 4.37e+06 & 1.87e+06 & 1.52e+06 & 1.40e+06\\
  10.00000 & 1.67e+08 & 1.61e+08 & 1.15e+08 & 6.87e+07 & 2.72e+07 & 1.30e+07 & 5.86e+06 & 5.08e+06 & 5.59e+06\\
  16.59590 & - & 4.91e+08 & 3.83e+08 & 2.20e+08 & 8.97e+07 & 4.55e+07 & 2.18e+07 & 1.86e+07 & 1.97e+07\\
  27.54230 & - & 1.80e+09 & 1.70e+09 & 8.98e+08 & 3.25e+08 & 1.70e+08 & 8.51e+07 & 6.80e+07 & 6.68e+07\\
  45.70880 & - & - & 4.97e+09 & 3.54e+09 & 1.23e+09 & 6.43e+08 & 3.31e+08 & 2.60e+08 & 2.39e+08\\
  75.85780 & - & - & - & 9.04e+09 & 4.69e+09 & 2.43e+09 & 1.30e+09 & 1.02e+09 & 8.98e+08\\
 125.89300 & - & - & - & 1.02e+10 & 1.03e+10 & 8.99e+09 & 5.10e+09 & 3.98e+09 & 3.41e+09\\
 208.92999 & - & - & - & - & 9.10e+09 & 1.05e+10 & 1.98e+10 & 1.51e+10 & 1.32e+10\\
 346.73700 & - & - & - & - & - & - & 9.19e+09 & 5.67e+10 & 5.14e+10\\
 575.44000 & - & - & - & - & - & - & - & 1.54e+10 & 1.30e+11\\

\enddata
\tablecomments{Column headers are temperatures in Kelvin.  Values are time constants in seconds.}
\label{tau189}
\end{deluxetable}

\end{document}